\PassOptionsToPackage{table,xcdraw}{xcolor}
\documentclass[conference]{IEEEtran}
\IEEEoverridecommandlockouts

\usepackage{tikz}
\usepackage{xcolor}
\usepackage{amsmath,amssymb,amsfonts}
\usepackage{graphicx}
\usepackage{tabularx}
\usepackage{multirow}
\usepackage{booktabs}
\usepackage{adjustbox}
\usepackage{circledsteps}
\usepackage{wasysym}
\usepackage{hyperref}
\usepackage{caption}
\usepackage{supertabular}
\usepackage{listings}
\usepackage{float}
\floatstyle{ruled} 
\newfloat{listing}{htbp}{lop}
\floatname{listing}{Listing}

\definecolor{mygreen}{rgb}{0,0.6,0} 

\lstdefinestyle{cppstyle}{
  language=C++,
  frame=lines,
  numbers=left,
  numberstyle=\tiny,
  xleftmargin=20pt,
  basicstyle=\ttfamily\tiny,
  keywordstyle=\color{purple}\bfseries,
  stringstyle=\color{mygreen},
  commentstyle=\color{teal}\itshape,
  showstringspaces=false,
  breaklines=true
}

\definecolor{light-gray}{gray}{0.95}

\newcolumntype{P}[1]{>{\centering\arraybackslash}p{#1}}

\definecolor{mycolor}{RGB}{9,99,125}

\newcommand{\bluesquare}[1]{%
    \tikz[baseline=(char.base)] 
        \node[fill=mycolor, rounded corners=2pt, minimum size=1em, inner sep=2pt, text=white] (char) {#1};%
}

\newcommand*\step[1]{\tikz[baseline=(char.base)]{
            \node[shape=circle,fill,inner sep=1pt,font=\scriptsize\bfseries] (char) {\textcolor{white}{#1}};}}

\begin{document}

\date{}

\title{\Large \bf A Practical Guideline and Taxonomy to LLVM's Control Flow Integrity}

\makeatletter
\newcommand{\linebreakand}{%
  \end{@IEEEauthorhalign}
  \hfill\mbox{}\par
  \mbox{}\hfill\begin{@IEEEauthorhalign}
}
\makeatother
\author{

\IEEEauthorblockN{Sabine Houy}
\IEEEauthorblockA{\textit{Ume\aa~University} \\
sabine.houy@umu.se}
\and
\IEEEauthorblockN{Bruno Kreyssig}
\IEEEauthorblockA{\textit{Ume\aa~University} \\
bruno.kreyssig@umu.se}
\and
\IEEEauthorblockN{Timoth\'ee Riom}
\IEEEauthorblockA{\textit{Ume\aa~University} \\
timothee.riom@umu.se}
\linebreakand
\IEEEauthorblockN{Alexandre Bartel}
\IEEEauthorblockA{\textit{Ume\aa~University} \\
alexandre.bartel@cs.umu.se}
\and
\IEEEauthorblockN{Patrick McDaniel}
\IEEEauthorblockA{\textit{University of Wisconsin-Madison} \\
 mcdaniel@cs.wisc.edu}
}

\maketitle

\begin{abstract}
Memory corruption vulnerabilities remain one of the most severe threats to software security. They often allow attackers to achieve arbitrary code execution by redirecting a vulnerable program's control flow. 
While Control Flow Integrity (CFI) has gained traction to mitigate this exploitation path, developers are not provided with any direction on how to apply CFI to real-world software.
In this work, we establish a taxonomy mapping LLVM’s forward-edge CFI variants to memory corruption vulnerability classes, offering actionable guidance for developers seeking to deploy CFI incrementally in existing codebases.
Based on the Top 10 Known Exploited Vulnerabilities (KEV) list, we identify four high-impact vulnerability categories and select one representative CVE for each. 
We evaluate LLVM’s CFI against each CVE and explain why CFI blocks exploitation in two cases while failing in the other two, illustrating its potential and current limitations. 
Our findings support informed deployment decisions and provide a foundation for improving the practical use of CFI in production systems.
\end{abstract}

\section{Introduction}\label{sec:introduction}
Memory corruption vulnerabilities remain one of the most critical threats to software security, consistently ranking among the top vulnerabilities in various threat lists. 
For instance, one specific form of memory corruption is ranked as the second most dangerous issue in the \emph{Most Dangerous Software Weaknesses} list~\cite{cweDangerous}, and memory corruption weaknesses occupy the top three spots in the \emph{Top 10 KEV\footnote{Known Exploited Vulnerabilities} Weaknesses}~\cite{cweKev}. 
These rankings underline the urgent need for robust and effective defenses.

Control-Flow Integrity (CFI) is a well-established defense mechanism designed to protect against control-flow hijacking attacks. 
However, despite its theoretical effectiveness, the real-world adoption of CFI remains inconsistent. 
Operating systems like Android and Windows have successfully integrated CFI~\cite{androidcfi,becker2024sok,houy2025twenty}, but Linux struggles with widespread adoption. 
Similarly, open-source browsers, including Chromium~\cite{chromiumcfi} and Firefox~\cite{firefoxcfi}, show limited support for CFI. 
One key challenge is that CFI can break functionality when policy violations occur, often triggered by factors such as C-standard mismatches or design decisions in legacy code. 
Additionally, while CFI research has produced various variants aimed at improving flexibility and efficiency~\cite{niu2014modular,niu2015per,burow2017control,bresch2020trustflow,jang2020ibv,park2023bgcfi,parisi2024titancfi,richter2024detrap}, there is still limited empirical data demonstrating how CFI performs against real-world exploits. 
This leaves developers uncertain about which CFI variant to deploy and how best to integrate it into existing projects.

The main CFI implementation applied in these cases is LLVM's; therefore, we focus on its seven CFI variants in this work. 
CFI was originally intended for the challenges faced by memory-unsafe languages, i.e., C and C++~\cite{abadi2009control}. 
LLVM also provides CFI for the ``unsafe'' code blocks in the otherwise memory-safe Rust language. 
While this highlights the ongoing relevance of CFI, in this work, we focus on the more imminent use case of memory-unsafe C/C++ code. 

The main contribution of this work is a systematic guideline that maps distinct classes of memory corruption vulnerabilities to the CFI variants most likely to provide practical protection. 
This guideline aims to support developers in making informed decisions about CFI integration, addressing the challenges posed by CFI’s incompatibility with existing software ecosystems~\cite{firefoxcfi,chromiumcfi,xu2019confirm,houy2024lessons,houy2025twenty}. 
To complement the guideline, we present four real-world exploits, each corresponding to a vulnerability type identified during its development. 
These examples help bridge the gap in the literature by showing where CFI mechanisms succeed or fail in practice, and offer insights into the reasons behind these outcomes. 
The examples are not meant to validate the guideline itself but serve to illustrate CFI’s practical strengths and limitations in diverse contexts.

We propose a structured guideline that maps different types of memory corruption vulnerabilities to LLVM’s CFI variants most likely to constrain control-flow manipulation. 
This mapping offers developers a practical reference for selecting CFI mechanisms that align with the specific vulnerabilities they aim to mitigate. 
To demonstrate the utility of the guideline, we select four recent high-impact CVEs\footnote{Common Vulnerabilities and Exposures} from different vulnerability classes. 
By analyzing these examples, we show how applying the appropriate CFI mechanisms could have prevented or significantly limited the exploitation of these vulnerabilities.

Our approach begins with a comprehensive review of prevalent memory corruption vulnerabilities to develop a structured taxonomy. 
This taxonomy serves as the foundation for mapping each vulnerability class to the CFI variants most likely to mitigate it. 
We then select four representative CVEs and compile the vulnerable software both with and without CFI enforcement. 
We test the publicly available PoCs/exploits against each compiled version to assess the impact of CFI on exploitability.

The real-world examples show that CFI mechanisms can successfully mitigate exploitation in some cases. 
In two out of four cases, CFI enforcement blocked exploit paths. 
In the remaining cases, CFI either partially restricted or failed to prevent exploitation, depending on the vulnerability type and CFI variant used. 
These examples underscore the need for developers to carefully consider the specific characteristics of the vulnerabilities they are targeting when applying CFI.

Main contributions:
\begin{enumerate}
    \item We present a guideline that systematically maps LLVM CFI variants to memory corruption vulnerability types, offering developers a structured approach to CFI deployment.
    \item We provide real-world examples that demonstrate when and why specific CFI variants succeed or fail in preventing exploitation. Additionally, we make the experimental environments and PoCs publicly available to facilitate reproducibility\footnote{Artifacts available at \url{https://github.com/software-engineering-and-security/cfi-practical-guideline.git}, including the resulting exploits and PoCs from our collection process (Section~\ref{sec:collection}).}.
\end{enumerate}

Our framework guides developers in selecting LLVM CFI variants that align with the specific memory corruption threats their applications face. 
By linking each variant to common vulnerability types, it helps prioritize protections based on application characteristics and deployment needs. 
Our findings support informed deployment decisions and provide a foundation for improving the practical use of CFI in production systems, especially by clarifying which variants are most relevant for mitigating specific classes of vulnerabilities.

While CFI significantly complicates exploitation, it remains, like all mitigation techniques~\cite{sacham2004aslr,snow2013just,butt2022depth}, ultimately bypassable~\cite{li2020finding,evans2015control,goktas2014out,conti2015losing,xu2023warpattack}. Throughout this work, \emph{prevent} refers to preventing specific exploit instances or triggers rather than achieving complete elimination of all vulnerability risks.

\section{Background}\label{sec:background}
\subsection{Control Flow Integrity (CFI)}\label{sec:backgroundCFI}
CFI is a fundamental defense mechanism against control-flow hijacking (CFH) attacks, such as Return-Oriented Programming (ROP) \cite{davi2014hardware, zhang2015control}, Jump-Oriented Programming (JOP) \cite{christoulakis2016hcfi}, and \texttt{vtable} corruption \cite{bounov2016protecting, LLVMDesigndoc}. 
CFI enforces that program control transfers adhere to a valid control flow graph (CFG), preventing attackers from redirecting execution to unintended code paths. 
Since many exploits rely on manipulating indirect control flow (e.g., function pointers, virtual table calls, type casting), various CFI variants exist to enforce integrity at different execution levels. 
CFI mechanisms are typically classified as either \emph{coarse-grained} or \emph{fine-grained}, depending on how strictly they restrict indirect control flow transfers. 
Coarse-grained CFI groups many legitimate targets into broad equivalence classes and allows indirect branching within that set. 
This approach is often vulnerable to advanced attacks that remain within these loose boundaries~\cite{carlini2015control}. 
In contrast, fine-grained CFI uses more precise static analysis to limit targets based on context, type information, or the position of the call site, significantly reducing the attack surface~\cite{tice2014enforcing}. 
As LLVM's CFI enforces type signature matching on the call side by verifying function prototypes, it is thus considered fine-grained~\cite{LLVMdoc,tice2014enforcing}.

\begin{figure}[ht!]
    \centering
    \includegraphics[width=0.9\linewidth]{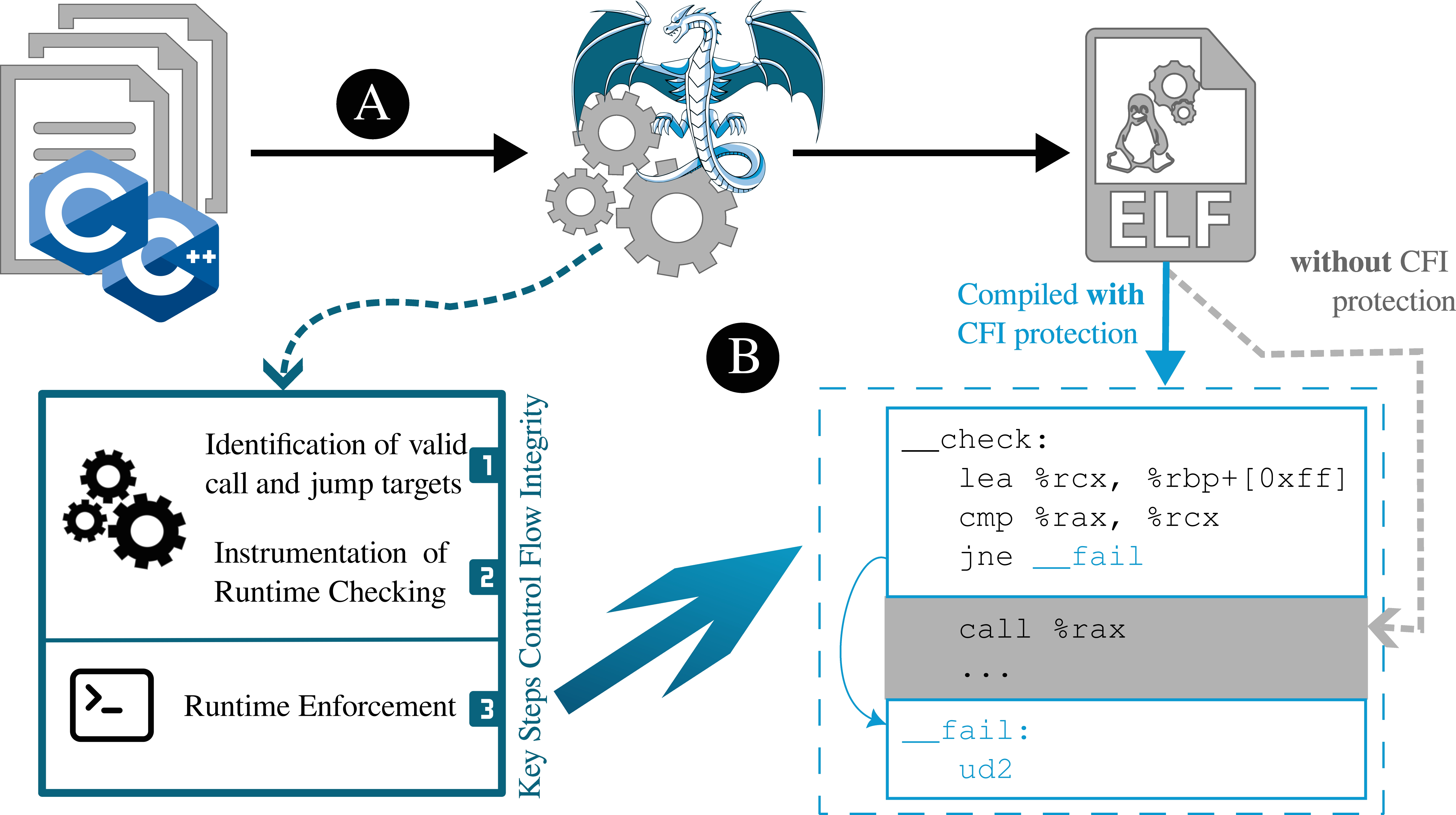}
    \caption{Control Flow Integrity simplified example}
    \label{fig:cfi_simple}
\end{figure}

Figure~\ref{fig:cfi_simple} illustrates the concept of CFI in a simplified example. 
Since CFI is a compiler-based mitigation technique, the compiler (in our case, LLVM \textsf{clang}/\textsf{clang++}~\cite{LLVMdoc}) introduces the CFI-related assembly instructions. 
This requires two steps (\Circled[inner color=white, fill color=black]{A}): 
First, the compiler identifies all valid call and jump targets (\bluesquare{\fontfamily{qag}\selectfont{1}}). 
Then, it generates the Control Flow Graph (CFG), determining all potential vulnerable jumps and calls. 
Subsequently, it creates a set of all \textbf{valid} call and jump targets, meaning an equivalence set of all targets with the same signature as the intended destination. 
LLVM uses a type-based signature relying on the return and parameter types. 
Then, for each potentially vulnerable jump and call, \textsf{clang} adds binary instructions validating the call target before the actual call is executed (\bluesquare{\fontfamily{qag}\selectfont{2}}). 
These additional checks are verified during runtime (\bluesquare{\fontfamily{qag}\selectfont{3}}). 

A simplified code example is shown in \Circled[inner color=white, fill color=black]{B}. 
Only the \texttt{call \%rax} instruction is part of the original code (no CFI enforcement). 
The code following the \texttt{\_\_check} and \texttt{\_\_fail} labels is the runtime enforcement, added in step (\bluesquare{\fontfamily{qag}\selectfont{2}}). 
Register \texttt{\%rax} contains the address of the to-be-performed call. 
CFI aims to verify that this address is valid within the CFG, meaning the signatures match. 
To ensure this, the predefined signature is loaded into register \texttt{\%rcx}. 
Then, the \texttt{\%rax}'s and \texttt{\%rcx}'s values are compared, determining whether the signatures are identical. 
If the comparison is successful, the call is executed, and the program's control flow proceeds as intended. 
Otherwise, the program jumps to \texttt{\_\_fail}, executing an undefined instruction (\texttt{ud2}) that leads to an \texttt{Illegal Instruction (SIGILL)} signal, causing the program to crash and preventing redirection to malicious code.

The CFG- and type-based CFI approaches have certain limitations. 
One key challenge is the \emph{over-approximation} of legitimate control-flow destinations. 
For instance, if a function to be protected has a \texttt{void*} return type and accepts an \texttt{int} as a parameter, any function in the CFG that matches this pattern will be considered a valid target. 
This can lead to a significant number of potential valid targets, which may allow an excessive number of jumps or function calls, reducing the effectiveness of CFI in preventing CFH attacks. 
In static analysis and program verification, over-approximation is recognized as an open research challenge~\cite{evans2015control,carlini2015control,tan2017cfg}. 

LLVM CFI combines CFG- and type-based implementation and offers four call-focused and three cast-focused variants, explained in~\ref{sec:call-focused} and~\ref{sec:cast-focused} respectively.

\subsubsection{CFI Variants for Function Call Integrity}\label{sec:call-focused}
\textbf{Non-virtual call CFI} (\texttt{cfi-nvcall}), ensures that function calls made through object pointers in non-virtual contexts cannot be redirected to unintended targets. 
Unlike indirect function calls, which resolve dynamically at runtime, non-virtual function calls are known at compile time. 
However, if an attacker corrupts an object pointer, execution may jump to arbitrary locations. 
Consider the following example depicted in Listing~\ref{lst:multi}: an attacker could manipulate \texttt{obj} (line 23) to point to a maliciously crafted memory region, causing \texttt{nonVirtualMethod()} to redirect the execution flow to unintended code. 
CFI verifies that \texttt{obj} remains a valid \texttt{NonVirtualClass} object before execution, preventing unauthorized function calls~\cite{LLVMdoc,tice2014enforcing,zhang2015control}.

Similarly, \textbf{virtual function call CFI} (\texttt{cfi-vcall}) ensures that virtual function calls always resolve to the correct method in a class's virtual table (\texttt{vtable}). 
C++ relies on \texttt{vtable}-based dynamic dispatch for polymorphism, making virtual function calls a prime target for attackers who overwrite \texttt{vtable} pointers~\cite{LLVMdoc,schuster2015counterfeit,davi2011ropdefender}.
\begin{listing}[H]
\begin{lstlisting}[style=cppstyle]
#include <iostream>

class BaseClass {
public:
    virtual void virtualMethod() {
        std::cout << "BaseClass::virtualMethod\n";
    }
};
class DerivedClass : public BaseClass {
public:
    void virtualMethod() override {
        std::cout << "DerivedClass::virtualMethod\n";
    }
};
class NonVirtualClass {
public:
    void nonVirtualMethod() {
        std::cout << "Legitimate function call\n";
    }
};
// cfi-nvcall
void callNonVirtualMethod(NonVirtualClass* obj) {
    obj->nonVirtualMethod();
}
// cfi-vcall
void callVirtualMethod(BaseClass* obj) {
    obj->virtualMethod();
}
\end{lstlisting}
\caption{C++ code example clarifying functionality of \texttt{cfi-nvcall} and \texttt{cfi-vcall}.}
\label{lst:multi}
\end{listing}
\vspace{-0.2cm}
In Listing~\ref{lst:multi} line 27, if \texttt{obj} 's virtual table pointer is overwritten, an attacker could redirect execution to an arbitrary function. 
CFI ensures that the virtual function call remains within the valid class hierarchy, mitigating \texttt{vtable} hijacking attacks~\cite{burow2017control,goktas2014out}. 
In other words, CFI would determine that the \texttt{virtualMethod} in \texttt{BaseClass} (line 5) and the overridden one in \texttt{DerivedClass} (line 11) are valid call targets within \texttt{obj}'s \texttt{vtable}.

Another common target is indirect function calls, which are protected by \textbf{indirect function call CFI} (\texttt{cfi-icall}). 
Indirect calls occur via function pointers, making them a prime vector for hijacking attacks~\cite{LLVMdoc,abadi2009control}. 
\begin{listing}[H]
\begin{lstlisting}[style=cppstyle]
void safeFunction() { 
    std::cout << "Safe function executed\n"; 
}
void callFunction(void (*func)()) {
    func();
}
int main() {
    void (*functionPointer)() = safeFunction;
    callFunction(functionPointer);
}
\end{lstlisting}

\caption{C++ code example clarifying functionality of \texttt{cfi-icall}.}
\label{lst:icall}
\end{listing}
\vspace{-0.2cm}
If an attacker overwrites \texttt{functionPointer} with a rogue address, shown in Listing~\ref{lst:icall}, execution may jump to malicious shellcode, ROP gadgets, or system functions. 
CFI restricts \texttt{functionPointer} to known valid function addresses within the program's CFG, blocking control flow hijacking~\cite{tice2014enforcing,zhang2015control}.

A related mechanism, \textbf{member-function call CFI} (\texttt{cfi-mfcall}), protects member function pointers in C++. 
Unlike ``\emph{regular}'' function pointers, member function pointers require both a valid object and a valid method, increasing the attack surface~\cite{LLVMdoc,zou2024tips,khandaker2019origin}.
\begin{listing}[H]
\begin{lstlisting}[style=cppstyle]
void callMemberFunction(BaseClass* obj, 
            void (BaseClass::*func)()) {
    (obj->*func)();
}
int main() {
        DerivedClass obj;
        void (BaseClass::*functionPointer)() = 
           &BaseClass::memberFunction;
        callMemberFunction(&obj, functionPointer);
}
\end{lstlisting}

\caption{C++ code example clarifying functionality of \texttt{cfi-mfcall}.}
\label{lst:mfcall}
\end{listing}
\vspace{-0.2cm}
If an attacker corrupts \texttt{functionPointer} or \texttt{obj}, see Listing~\ref{lst:mfcall}, execution may be redirected to arbitrary memory locations. CFI ensures that \texttt{functionPointer} remains within the correct class hierarchy, preventing exploits that target \texttt{vtable} corruption.

While both \texttt{cfi-icall} and \texttt{cfi-nvcall} enforce function call integrity, they protect against distinct attack vectors. 
\texttt{cfi-icall} applies to indirect function calls via function pointers. 
Since function pointers resolve dynamically, attackers can overwrite them to execute arbitrary code. 
CFI ensures that only legitimate function addresses are called, blocking hijacking attempts. 
\texttt{cfi-nvcall} applies to non-virtual function calls, typically resolved at compile time. 
However, if an attacker corrupts an object pointer, the function call could target unintended memory regions. 
CFI prevents this by ensuring the object pointer remains valid in its expected class.

\subsubsection{CFI Variants for Type Casting Protection}\label{sec:cast-focused}

Beyond function calls, CFI also enforces type-casting constraints to prevent type confusion and object corruption attacks. 
The \texttt{cfi-unrelated-cast} variant ensures that \textbf{casts between unrelated types} are prohibited, preventing memory corruption vulnerabilities which affect the control flow~\cite{LLVMdoc,schuster2015counterfeit}.
\begin{listing}[H]
\begin{lstlisting}[style=cppstyle]
int number = 42;
float* invalidPointer = 
    reinterpret_cast<float*>(&number);
\end{lstlisting}

\caption{C++ code example clarifying functionality of \texttt{cfi-unrelated-cast}.}
\label{lst:unrelated-cast}
\end{listing}
\vspace{-0.2cm}
If an attacker misuses \texttt{reinterpret\_cast}, they may introduce data corruption or arbitrary code execution. 
CFI blocks such casts at runtime, ensuring type safety. 

In contrast, \texttt{cfi-derived-cast} ensures that \textbf{casting between \emph{base} and \emph{derived} classes} remains valid, mitigating type confusion attacks~\cite{LLVMdoc,zhang2015control}.
\begin{listing}[H]
\begin{lstlisting}[style=cppstyle]
void testCast(BaseClass* basePtr) {
    DerivedClass* derivedPtr = 
        dynamic_cast<DerivedClass*>(basePtr);
}
\end{lstlisting}

\caption{C++ code example clarifying functionality of \texttt{cfi-derived-cast}.}
\label{lst:derived-cast}
\end{listing}
\vspace{-0.2cm}
An unsafe cast may occur if \texttt{basePtr} is manipulated to point to an unrelated object; see Listing~\ref{lst:derived-cast}.
CFI prevents such abuses by validating inheritance relationships. 
A stricter enforcement mechanism, \texttt{cfi-cast-strict}, extends \texttt{cfi-derived-cast} by covering cases where a derived class introduces no new virtual functions and has an identical memory layout to its base class~\cite{burow2017control}.

\subsection{CFI Relevant Memory Corruption Vulnerabilities}
CFI is specifically designed to prevent CFH attacks. 
The most relevant vulnerabilities to CFI are those that allow attackers to directly divert the program's control flow, typically by corrupting code pointers, return addresses, or function calls. 
There are many different types and subcategories of memory corruption vulnerabilities. 
In general, CFI can only protect against attacks that lead to a modified control flow. 
We choose the vulnerabilities based on the current Top 10 KEV list, ranking \emph{use-after-free} and \emph{heap-based buffer overflows} first and second, respectively. 
Place three is ``Out-of-bounds Write'', the heap- and stack-based buffer overflows parent category, thus adding \emph{stack-based buffer overflows}. 
\emph{Type confusion} reaches rank number eight. 
All other listed vulnerabilities cannot directly corrupt the memory (e.g, Server-Side Request Forgery -SSRF) and are thus not directly relevant to CFI and the scope of this paper. 
However, they could affect the memory by triggering, for instance, a buffer overflow, which would be covered by \emph{heap- stack-based buffer overflows}. 

\subsubsection{Stack-based Buffer Overflow} 
A stack-based buffer overflow is a common vulnerability in C and C++ programs. 
It arises when data is written beyond the boundaries of a buffer allocated on the stack. 
This issue often occurs when functions such as \texttt{strcpy}, \texttt{sprintf}, or \texttt{gets} are used without proper bounds checking, enabling more data to be written than the buffer can accommodate. 
Consequently, this excess data can overwrite adjacent memory on the stack, including critical control information, e.g., return addresses, function pointers, and local variables~\cite{one1996smashing,wagle2003stackguard,bhatkar2003address}. 
This leads to unpredictable behavior, crashes, or, in some cases, exploitation by attackers who deliberately craft inputs to overwrite control structures, redirecting execution to malicious code. 
Stack-based buffer overflows are particularly dangerous because they can be exploited for arbitrary code execution, privilege escalation, or denial-of-service attacks~\cite{wei2016survey}. 

Modern compiler protections and operating systems offer enhanced defense mechanisms, including stack canaries, Address Space Layout Randomization (ASLR), and non-executable stacks. 
However, as all these techniques are bypassable in different ways~\cite{richarte2002four,bierbaumer2018smashing,jaloyan2020return,evtyushkin2016jump}, it is essential to use multiple at once and add as many security layers as possible. 
An additional layer of security, such as enforcing CFI, can improve resistance to stack-based buffer overflows~\cite{cowan1998stackguard,wagle2003stackguard,wei2016survey}.

\subsubsection{Heap-based Buffer Overflow} 
A heap-based buffer overflow occurs when a buffer allocated in the heap overflows, resulting in the overwriting of adjacent memory regions. 
This can corrupt heap structures, alter program data, or facilitate arbitrary code execution. 
A successful attack may overwrite return addresses or critical metadata, allowing attackers to redirect control flow. 
Heap-based buffer overflows typically arise when functions like \texttt{malloc}, \texttt{calloc}, or \texttt{realloc} are used to allocate memory, and the data is written into the buffer without adequately checking its size. 
This oversight can cause the buffer to overflow, potentially corrupting adjacent heap structures or metadata used by the memory allocator. 
Attackers can exploit these vulnerabilities to manipulate the heap layout, overwrite critical data, such as function pointers or object metadata, and achieve arbitrary code execution or privilege escalation~\cite{zeng2015heaptherapy,jia2017towards,butt2022depth}. 

Additionally, runtime protections like heap hardening mechanisms, memory sanitizers, and allocators with built-in integrity checks, such as LLVM's CFI, offer necessary safeguards~\cite{jia2017towards}.

\subsubsection{Use-After-Free (UAF)} 
A use-after-free (UAF) vulnerability arises when a program continues to access memory after it has been freed, resulting in undefined behavior. 
This allows attackers to manipulate freed memory for malicious purposes, such as executing arbitrary code or causing crashes. 
UAF vulnerabilities are particularly prevalent in languages like C and C++, where memory management is manual. 
These vulnerabilities occur when a pointer continues to reference a memory location after it has been freed using \texttt{free} in C or \texttt{delete} in C++. 
If the pointer is not nullified, it becomes a ``dangling pointer'', which can be dereferenced to access reallocated or corrupted memory, leading to potential exploits~\cite{gui2021automated,lu2022survey,chen2023all}. 

Techniques like CFI can help detect and prevent such vulnerabilities by ensuring valid control flow paths.

\subsubsection{Type Confusion} 
A type confusion vulnerability occurs when a program mistakenly treats an object in memory as a different type than it is, often leading to unsafe memory operations. 
This can result in undefined behavior, program crashes, or allow attackers to execute arbitrary code. 
Type confusion vulnerabilities are common in object-oriented programming and commonly lead to type mismatches in virtual tables (\texttt{vtable}), altering the intended control flow. 
In C and C++, type confusion typically arises when an object of one type is misinterpreted as another, often due to improper type casting or unsafe memory manipulation. 
For example, if a pointer to an object of type \texttt{A} is cast to a pointer of type \texttt{B}, the program may misinterpret the underlying memory, resulting in undefined behavior or security breaches~\cite{jeon2017hextype,zou2019tcd}. 
Attackers can exploit this vulnerability to corrupt memory, bypass security checks, or gain control of the program’s execution path.

Static analyzers, runtime type verification tools, and techniques like Control Flow Integrity are also critical in ensuring that type assumptions are correctly enforced.

\section{Implementation of Systematization}\label{sec:methodology}
In the following, we discuss to what extent LLVM's CFI implementation can potentially mitigate the mentioned vulnerabilities (described in detail in Section~\ref{sec:background}). 

CFI protects against specific types of \emph{buffer overflows} if data related to the control flow is corrupted. 
If a heap or stack corruption modifies data fields unrelated to control flow,  CFI cannot detect the exploit when no protected function or \texttt{vtable} pointers are corrupted on the way. 
In the case of \emph{use-after-free (UAF)}, CFI can mitigate vulnerabilities if a freed object’s \texttt{vtable} or function pointers are reused inappropriately. 
CFI can effectively only prevent \emph{type confusion} from malicious casts or directly altering function pointers, thus resulting in a mismatched type.  

To summarize, CFI checks are effective against memory corruptions that involve altering function pointers or \texttt{vtables}. 
In the following section, we elaborate on how LLVM's implementation of CFI can affect the four top 10 KEV vulnerabilities.

\subsection{Mapping of Relevant Memory Corruption Vulnerabilities to CFI Types}\label{sec:mapping}

The following explains why specific CFI variants are tailored to specific threat models (types of memory corruption vulnerabilities) rather than being universal solutions. 
We provide a mapping for each vulnerability category and its expected effective CFI variants~\cite{abadi2009control,tice2014enforcing,LLVMdoc} (see Section~\ref{sec:backgroundCFI}), summarized in Table~\ref{tab:mapping}. 

Our rating system uses a three-point scale: low (\Circle), moderate (\RIGHTcircle), and high (\CIRCLE) mitigation potential. 
It is important to note that \textit{high} does not mean that CFI will protect against all exploits based on this vulnerability type. 
Rather, it indicates which CFI variants most effectively block common exploitation vectors of the respective vulnerabilities. 
The overall score for each vulnerability is calculated as the mean of all CFI variants. 

Each variant is assigned one of these levels based on an approximation based on a summary of facts extracted from existing scientific literature.  
We performed a keyword search (based on the considered vulnerabilities and CFI) in 2024 in major scientific library databases (such as ACM and IEEE) to find relevant literature. 
Then, we filtered based on the title and then on the abstract. 
Thus, literature focusing solely on other languages, such as Java, would be excluded. 
When identifying relevant work, we performed backward- and forward-referencing to find additional literature. 
\begin{table}[htb!]
\centering
    \caption{Literature used for Mapping}
    \label{tab:literature}
\begin{adjustbox}{width=0.9\linewidth}
    \begin{tabular}{rp{5cm}}
        \toprule
        
        \multicolumn{1}{r}{\begin{tabular}[r]{@{}r@{}}
            \textsc{Vulnerability} \\
            \textsc{Type}
        \end{tabular}}                                  &
        \textsc{Literature}                             \\
        \midrule

        \rowcolor{light-gray}
        \multicolumn{1}{r}{\begin{tabular}[r]{@{}r@{}}
            Heap-based          \\
            Buffer Overflow
        \end{tabular}}                                  &
        \cite{burow2017control,conti2015losing,sayeed2019control,abadi2009control,jia2017towards,cowan1998stackguard,zeng2015heaptherapy,tice2014enforcing,ammar2024bridging,goktas2014out,LLVMdoc,heapbased,bufferoverflow}                                                \\
        
        \multicolumn{1}{r}{\begin{tabular}[r]{@{}r@{}}
            Stack-based         \\
            Buffer Overflow
        \end{tabular}}                                  & 
        \cite{cowan1998stackguard,nicula2019exploiting,padaryan2015automated,xu2022bofaeg,abadi2009control,tice2014enforcing,sayeed2019control,burow2017control,bierbaumer2018smashing,zhang2013practical,ammar2024bridging,maunero2019cfi,LLVMdoc,stackbased,bufferoverflow}                                                \\

        \rowcolor{light-gray}                                                
        \multicolumn{1}{r}{\begin{tabular}[r]{@{}r@{}}
            Use-After-Free      \\
            (UAF)               
        \end{tabular}}                                  &
        \cite{ammar2024bridging,conti2015losing,lee2015preventing,sayeed2019control,abadi2009control,tice2014enforcing,jia2017towards,zeng2015heaptherapy,LLVMdoc,uaf,owasp-uaf,li2020finding}                                                \\

        \multicolumn{1}{r}{\begin{tabular}[r]{@{}r@{}}
            Type                \\
            Confusion  
        \end{tabular}}                                  & 
        \cite{badoux2025type++,LLVMdoc,kim2024bintyper,fan2023accelerating,farkhani2018effectiveness,jeon2017hextype,li2020finding,muntean2018castsan,typeconf,pang2018mapping,zou2019tcd,cowan1998stackguard,tice2014enforcing}                                                \\
        
        \bottomrule                                         \\
    \end{tabular}
    \end{adjustbox}
    \vspace{-0.5cm}
\end{table}

The resulting literature is depicted in Table~\ref{tab:literature}.
In addition, we provide a code book containing direct quotes of the relevant literature used to create the mapping, in the Appendix Table \ref{tab:codebook}. 
The extracted quotes do not mean that they are the only relevant ones, but the most relevant ones.

\begin{table*}[ht!]
\centering
    \caption{Mapping of Memory Corruption Vulnerabilities to LLVM's CFI Mechanisms}
    \label{tab:mapping}
	\begin{adjustbox}{width=0.8\linewidth}
		\begin{tabular}{@{}rcccc@{}}
			\toprule
			\multicolumn{1}{c}{}                                            
			& \multicolumn{4}{c}{\textbf{Memory Corruption Vulnerability}}
			\\
			\multicolumn{1}{l}{\textbf{CFI Variants}}&&&&\\
			
			& \multicolumn{1}{r}{\begin{tabular}[c]{@{}c@{}}\textsc{Heap-based}\\ \textsc{Buffer Overflow}\end{tabular}}
			& \multicolumn{1}{r}{\begin{tabular}[c]{@{}c@{}}\textsc{Stack-based}\\ \textsc{Buffer Overflow}\end{tabular}}
			& \multicolumn{1}{r}{\begin{tabular}[c]{@{}c@{}}\textsc{Use-After}\\ \textsc{-Free (UAF)}\end{tabular}}
			& \multicolumn{1}{r}{\begin{tabular}[c]{@{}c@{}}\textsc{Type}\\ \textsc{Confusion}\end{tabular}} 
			\\ \midrule

			\rowcolor{light-gray}
			\texttt{cfi-cast-strict}			             &
			${\RIGHTcircle}^{\textcolor{white}{*}}$      &
			${\Circle}^{*}$					             &
			${\RIGHTcircle}^{\textcolor{white}{*}}$      &
			${\CIRCLE}^{\textcolor{white}{*}}$           \\
			
			\texttt{cfi-derived-cast}                       & 
			${\RIGHTcircle}^{*}$                         &
			${\Circle}^{*}$                              &
			${\Circle}^{*}$                              &
			${\CIRCLE}^{\textcolor{white}{*}}$           \\
			
			\rowcolor{light-gray}
			\texttt{cfi-unrelated-cast}                     & 
			${\RIGHTcircle}^{*}$                         &			
			${\Circle}^{*}$                              &
			${\Circle}^{*}$                              &
			${\CIRCLE}^{\textcolor{white}{*}}$           \\
			
			\texttt{cfi-nvcall}                   & 
			${\Circle}^{*}$                              & 			
			${\RIGHTcircle}^{\textcolor{white}{*}}$      & 
			${\Circle}^{\textcolor{white}{*}}$           & 
			${\Circle}^{*}$                              \\
			
			\rowcolor{light-gray}
			\texttt{cfi-vcall}                       & 
			${\CIRCLE}^{\textcolor{white}{*}}$           &
			${\Circle}^{\textcolor{white}{*}}$           &
			${\RIGHTcircle}^{\textcolor{white}{*}}$      &
			${\RIGHTcircle}^{*}$                         \\
			
			\texttt{cfi-icall}                      & 
			${\CIRCLE}^{\textcolor{white}{*}}$           & 
			${\CIRCLE}^{\textcolor{white}{*}}$           & 
			${\RIGHTcircle}^{\textcolor{white}{*}}$      & 
			${\Circle}^{*}$                              \\
			
			\rowcolor{light-gray}
			\texttt{cfi-mfcall}               & 
			${\Circle}^{*}$                              & 
			${\Circle}^{*}$                              & 
			${\Circle}^{*}$                              & 
			${\Circle}^{*}$                              \\ 
			
			\midrule
			\begin{tabular}[c]{@{}c@{}}\textbf{Average}\\ \textbf{\emph{mitigation potential}}\end{tabular}         &
			\emph{moderate}                  &
			\emph{low}                       &
			\emph{low}                       &
			\emph{moderate}                  \\
			
			\bottomrule
		\end{tabular}
	   \end{adjustbox}
        \begin{center}
            \textbf{Legend:} \Circle~= \emph{low} , \RIGHTcircle~= \emph{moderate}, \CIRCLE~= \emph{high} mitigation potential\\
            ${}^{*}$on entries where the CFI variant was \underline{not intended} to mitigate the respective vulnerability type.
        \end{center}
        \vspace{-0.4cm}
\end{table*}

\subsubsection{Heap-based Buffer Overflow}
Heap overflows can overwrite function pointers, vtable pointers, and other heap-resident control-flow structures. As several works highlight forward-edge hijacks, LLVM's \texttt{icall}\footnote{We omit \texttt{cfi-} in the following for readability.} and \texttt{vcall} CFI are particularly relevant.

\CIRCLE~\emph{High} – \texttt{icall}, \texttt{vcall}:
Tice et al.~\cite{tice2014enforcing} and Sayeed et al.~\cite{sayeed2019control} explicitly describe heap corruption of indirect call targets and \texttt{vtable} pointers, both of which are precisely the targets protected by \texttt{icall} and \texttt{vcall} CFI. 
Abadi et al.~\cite{abadi2009control} also show CFI’s effectiveness in preventing jump-to-libc and ROP-based heap attacks.

\RIGHTcircle~\emph{Medium} – \texttt{cast-strict}, \texttt{derived-cast}, \texttt{unrelated-cast}:
Literature, e.g., Abadi et al.~\cite{abadi2009control} and Burow et al.~\cite{burow2017control}, link heap overflows to type confusion and control-flow subversion. 
These cast-based protections enforce type soundness, helping mitigate, but not fully prevent, such attacks. 
Sayeed et al. mention IFCC protecting ``\emph{forward-edge}" heap resident data, which aligns partially with cast variants.

\Circle~\emph{Low} – the variants \texttt{nvcall}, \texttt{mfcall}
 were not directly implicated in any reviewed heap-based buffer overflow scenarios. 
 The literature does not indicate widespread use of heap corruption to subvert static member functions or non-virtual calls.

\subsubsection{Stack-based Buffer Overflow}
Stack overflows have long targeted return addresses and now commonly enable Return-Oriented Programming (ROP). Although CFI aims to protect backward edges, return instructions are typically outside the scope of LLVM's standard CFI variants.

\CIRCLE~\emph{High} – \texttt{icall}:
Abadi et al.~\cite{abadi2009control}, Burow et al.~\cite{burow2017control}, and Padaryan et al.~\cite{padaryan2015automated} describe how stack smashing can affect function pointers stored in stack memory. Protecting such pointers falls under the domain of \texttt{icall} if used in indirect calls.

\RIGHTcircle~\emph{Medium} – \texttt{cast-strict}:
Abadi et al.~\cite{abadi2009control} describe how CFI prevents redirection via corrupted function pointers (even when originating from stack misuse), offering limited coverage depending on pointer usage. \texttt{cast-strict} offers partial defense if the vulnerability chain includes illegitimate pointer casting.

\Circle~\emph{Low} – all others:
Most variants are not relevant in typical stack-based attacks. Stack-based buffer overflows typically do not involve \texttt{vtables}, dynamic casts, or member-function pointers, leading to low mitigation relevance.

\subsubsection{Use-After-Free (UAF)} 
UAF bugs can reuse dangling pointers to overwrite or fake object layouts—especially \texttt{vtable} or function pointers—making CFI for dynamic call targets essential.

\RIGHTcircle~\emph{Medium} – \texttt{vcall}, \texttt{cast-strict}, \texttt{derived-cast}:
Tice et al.~\cite{tice2014enforcing} describe how attackers reuse freed objects to inject malicious \texttt{vtables}, making \texttt{vcall} CFI highly relevant. 
However, Lee et al.~\cite{lee2015preventing} and Sayeed et al.~\cite{sayeed2019control} note that cast-based defenses help restrict control flow in UAF exploitation, but attackers may still perform non-control-data attacks. 
Badoux et al.~\cite{badoux2025type++} and Yuan Li et al.~\cite{li2020finding} describe the use of type-enforcing CFI (like LLVM's cast protections) to mitigate attacks relying on type confusion induced by UAF. 

\Circle~\emph{Low} – all others:
Despite their use in dispatching calls, \texttt{icall}, \texttt{nvcall}, and \texttt{mfcall}, these variants were not directly associated with UAF attack mechanisms in the literature. 
While function and member-function pointers might be targets in UAF, no sources documented exploits using them via freed memory.

\subsubsection{Type Confusion} 
Type confusion occurs when an object is misinterpreted as another type, often via unsafe casting. This can lead to arbitrary memory access, misused \texttt{vtables}, or invalid calls—making type- and callsite-aware CFI essential.

\CIRCLE~\emph{High} – \texttt{cast-strict}, \texttt{unrelated-cast}:
Multiple papers (e.g., Farkhani et al.~\cite{farkhani2018effectiveness}, Pang et al.~\cite{pang2018mapping}, Fan et al.~~\cite{fan2023accelerating}) describe runtime type enforcement (RTC-based CFI) as the most effective strategy to prevent control-flow hijacks via type confusion. 
These works directly support high mitigation potential for LLVM’s \texttt{cast-strict} and \texttt{unrelated-cast}, both of which enforce type compatibility at runtime. 
Badoux et al.~\cite{badoux2025type++} explain that LLVM-CFI targets polymorphic casts, mapping closely to these variants.

\RIGHTcircle~\emph{Medium} – \texttt{derived-cast}:
Unsafe downcasts are a common source of confusion, for instance, Kim and Kim~\cite{kim2024bintyper}, and \texttt{derived-cast} specifically protects against misuse of base-to-subclass conversions. However, it does not address all cases, such as unrelated casts or deep type violations.

\Circle~\emph{Low} – \texttt{vcall}, \texttt{nvcall}, \texttt{icall}, \texttt{mfcall}: 
Although type confusion can lead to hijacked call targets, e.g., forged \texttt{vtables}, the function-call variants only mitigate attacks when the call itself is reached. 
They do not enforce type correctness at the point of cast, making them insufficient against the root cause of confusion-based exploits.

\section{Application \& CVE Example Evaluation}\label{sec:results}
\subsection{CVE Collection Process}\label{sec:collection}
We systematically filter and collect Common Vulnerabilities and Exposures (CVEs) to identify relevant and potentially exploitable vulnerabilities. 

\begin{figure}[ht!]
    \centering
    \includegraphics[width=\linewidth]{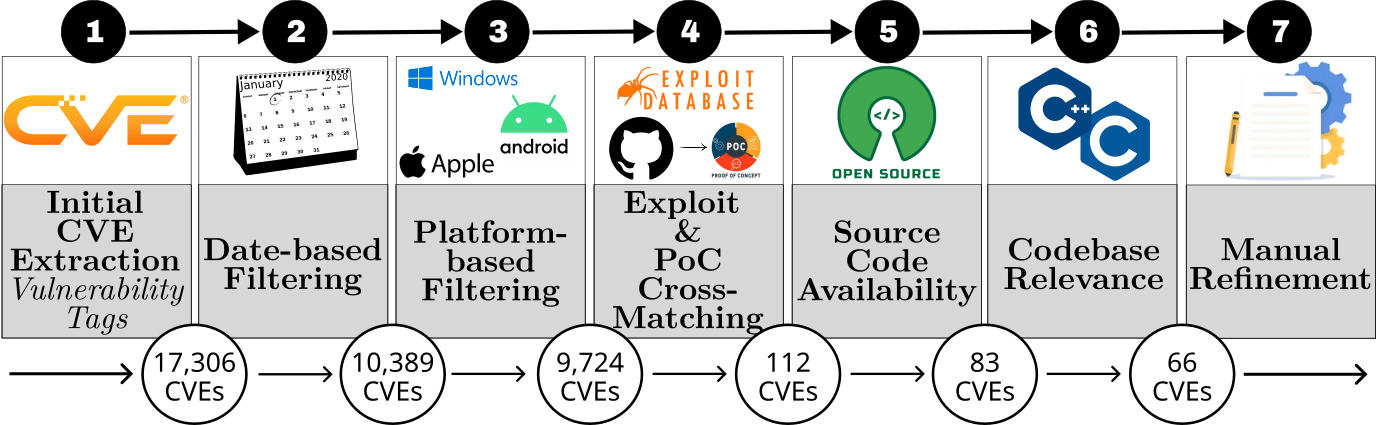}
    \caption{CVE Collection Process}
    \label{fig:collection_process}
\end{figure} 
Figure~\ref{fig:collection_process} illustrates these steps.
In the first step, \step{1}-\textbf{Initial CVE Extraction}, we gather all CVEs from the MITRE~\cite{mitrecve} database that are tagged with vulnerability types pertinent to our research as discussed in Section~\ref{sec:mapping} and based on the \emph{Top 10 KEV\footnote{Known Exploited Vulnerabilities} Weaknesses} list~\cite{cweKev}. 
Next, we use \step{2}-\textbf{Date-based Filtering} to exclude CVE entries that predate the year 2018. 
This ensures focus on contemporary security challenges and, more importantly, to excludes CVEs before the introduction of LLVM's CFI in \textsf{Clang 3.7}~\cite{llvmcfiintro}.
It additionally decreases incompatibilities with older C/C++ standards~\cite{oldstandards1,oldstandards2}. 
Following this, \step{3}-\textbf{Platform-based Filtering} is performed to eliminate CVEs that exclusively affect Windows, Android, MacOS, or iOS.
Since we focus on vulnerabilities impacting open-source software ecosystems, MacOS and iOS are irrelevant. 
These operating systems implement their own CFI mechanisms. 
Windows~\cite{windowscfi} and Android~\cite{androidcfi} fork the implementation of LLVM and add their features, whereas MacOS~\cite{applecfi} uses Pointer Authentication Codes, a separate LLVM mitigation technique~\cite{llvmpac}. 
These aspects place all of them outside the scope of our study.
The fourth step, \step{4}-\textbf{Exploit \& PoC Cross-Matching}, involves identifying CVEs with publicly available exploits by cross-referencing ExploitDB~\cite{explloitdb} and extensive PoC repositories on GitHub~\cite{pocrepository}. 
As shown in Figure~\ref{fig:collection_process}, this step led to the removal of the highest number of CVEs. 
We then apply \step{5}-\textbf{Source Code Availability} filtering, removing all CVEs not pertaining to open-source software, as access to source code is necessary for further security assessments. 
Next, \step{6}-\textbf{Codebase Relevance} filtering ensures that only C or C++ code vulnerabilities are retained, as CFI applies to these languages.

Finally, \step{7}-\textbf{Manual Refinement} ensures that CVEs related to Just-in-Time (JIT) compiled code are excluded, as CFI cannot effectively mitigate threats in such dynamically generated environments~\cite{niu2014rockjit,bauman2023renewable}. 
This results in excluding almost all type confusion vulnerabilities identified in web browsers such as Firefox and Chrome. 
Most type confusion vulnerabilities are found in the JavaScript engine of the browser (Firefox's SpiderMonkey~\cite{spidermonkey} and Chrome's V8~\cite{v8}) and rely heavily on JITed code, resulting in excluding an additional 28 CVEs. 
Current implementations of CFI cannot protect it since the JITed code is unavailable during compilation. 
Moreover, we also omit all vulnerabilities identified within the Linux kernel as LLVM provides a separate implementation of CFI for the kernel (\texttt{kcfi})~\cite{kcfillvm,kcfi} (removing five CVEs), which is out-of-scope for this work. 

This structured multi-step process identifies only the most relevant and applicable CVEs. 
While this process strongly limits the amount of real-world exploits we can find, it is necessary for targeted research and analysis.

\subsection{Experiment Setup}

For each CVE, we set up a \textsf{podman}~\cite{podman} container with the most compatible OS and version. 
We compile the vulnerable software without CFI using \textsf{clang}/\textsf{clang++} to confirm that the PoC or exploit functions as expected. 
This serves as the baseline for comparison with builds using each of the seven CFI variants. 
If behavior diverges, we recompile only the vulnerable function with CFI to confirm the violation. 
We also adjust build configurations as needed for \textsf{clang} and CFI compatibility.

\subsection{\emph{Example} of Guideline Application}
The specific security requirements and prioritization strategies inherently depend on the characteristics of the target project and the associated threat model.
In this section, we present a representative use case scenario for our taxonomy.

Suppose a developer wants to improve the overall security of an application by integrating CFI mechanisms.
Based on Table~\ref{tab:mapping}, a first step could be to deploy \texttt{icall}, as it has a high impact (\CIRCLE) in mitigating heap and stack-based buffer overflows, while only having a low score (\Circle) for type confusion vulnerabilities.

After deployment, two outcomes are possible: (i) the application remains functionally correct and allows the gradual introduction of additional CFI protections, or (ii) runtime errors occur that require debugging or re-evaluating the approach.
In both cases, the developer must make informed decisions about the subsequent integration steps.
If the application is prone to type confusion, it is recommended to prioritize the implementation of CFI cast variants.

The proposed methodology is in line with practices used by mature projects such as Chromium, which emphasizes the protection of indirect and virtual function calls as a key strategy to increase security~\cite{chromiumcfi}.

\subsection{Evaluation of CVEs with CFI}
\begin{table}[htb!]
\centering
    \caption{Example CVEs}
    \label{tab:cves}
\begin{adjustbox}{width=\linewidth}
    \begin{tabular}{rrccc}
        \toprule
        
        \multicolumn{1}{r}{\textsc{CVE ID}}                                     & 
        \multicolumn{1}{c}{\begin{tabular}[c]{@{}c@{}}
            \textsc{Affected}   \\ 
            \textsc{Project} 
        \end{tabular}}                                      & 
        \multicolumn{1}{c}{\begin{tabular}[c]{@{}c@{}}
            \textsc{Affected}   \\ 
            \textsc{Version} 
        \end{tabular}}                                      & 
        \textsc{PoC/Exploit}                                & 
        \textsc{Severity}                                   \\
        
        \midrule

        \rowcolor{light-gray}
        \multicolumn{1}{r}{\begin{tabular}[r]{@{}r@{}}\
            \emph{CVE-2021-3156}              \\
            \cite{CVE-2021-3156}  
        \end{tabular}}                                      & 
        \multicolumn{1}{r}{\begin{tabular}[r]{@{}r@{}}
            \textsf{sudo}                     \\
            \cite{sudo}      
        \end{tabular}}                                      & 
        \texttt{\textless{}1.9.5p2}                         & 
        \multicolumn{1}{c}{\begin{tabular}[c]{@{}c@{}}
            \textsc{Heap Overflow\text{*}}             \\
            Privilege escalation               \\ 
            \cite{exploit-CVE-2021-3156}
        \end{tabular}}                                      & 
        7.8 (High)                                          \\
        
        \multicolumn{1}{r}{\begin{tabular}[r]{@{}r@{}}
            \emph{CVE-2023-49992}               \\
            \cite{CVE-2023-49992}
        \end{tabular}}                                      & 
        \multicolumn{1}{r}{\begin{tabular}[r]{@{}r@{}}
            \textsf{espeak-NG}                  \\
            \cite{espeak-ng}  
        \end{tabular}}                                      & 
        \texttt{1.52-dev}                                   & 
        \multicolumn{1}{c}{\begin{tabular}[c]{@{}c@{}}
            \textsc{Stack Overflow\text{*}}             \\
            corrupting the stack                \\
            \cite{exploit-CVE-2023-49992} 
        \end{tabular}}                                      & 
        \multicolumn{1}{c}{\begin{tabular}[c]{@{}c@{}}
            5.3 (Medium)                        \\ 
            7.8 (High)\cite{ghsa-espeak}
        \end{tabular}}                                      \\

        \rowcolor{light-gray}
        \multicolumn{1}{r}{\begin{tabular}[r]{@{}r@{}}
            \emph{CVE-2022-3666}                \\
            \cite{CVE-2022-3666}  
        \end{tabular}}                                      & 
        \multicolumn{1}{r}{\begin{tabular}[r]{@{}r@{}}
            \textsf{Bento4}                     \\
            \cite{bento4}  
        \end{tabular}}                                      & 
        \texttt{1.6.0-639}                                  & 
        \multicolumn{1}{c}{\begin{tabular}[c]{@{}c@{}}
            \textsc{UAF}                        \\
            double-free                         \\
            \cite{exploit-CVE-2022-3666} 
        \end{tabular}}                                      & 
        7.8 (High)                                          \\
        
        \multicolumn{1}{r}{\begin{tabular}[r]{@{}r@{}}
            \emph{CVE-2024-34391}               \\
            \cite{CVE-2024-34391}  
        \end{tabular}}                                      &
        \multicolumn{1}{r}{\begin{tabular}[r]{@{}r@{}}
            \textsf{libxmljs}           \\
            \cite{libxmljs} 
        \end{tabular}}                                      & 
        \texttt{\textless{}=1.0.11}                        & 
        \multicolumn{1}{c}{\begin{tabular}[c]{@{}c@{}}
            \textsc{Type Confusion}     \\
            type misinterpretation      \\
            \cite{exploit-CVE-2024-34391} 
        \end{tabular}}                                      & 
        \multicolumn{1}{c}{\begin{tabular}[c]{@{}c@{}}
            N/A                         \\ 
            9.2 (Critical)              \\
            \cite{ghsa-libxmljs}     
        \end{tabular}}                                      \\
        
        \bottomrule                                         \\
    \end{tabular}
    \end{adjustbox}
    \small\text{*}refers to heap- and stack-based buffer overflow respectively.
    \vspace{-0.4cm}
\end{table}
We identified four relevant CVEs matching all necessary requirements described in Section~\ref{sec:collection}.
An overview of these CVEs is displayed in Table~\ref{tab:cves}. 
None of the projects containing the vulnerabilities (CVEs) support CFI by default. 
\begin{table*}[htb!]
\centering
    \caption{Results of CVEs compiled with LLVM's CFI Mechanisms}
    \label{tab:results}
	\begin{adjustbox}{width=0.8\linewidth}
		\begin{tabular}{@{}rcccc@{}}
			\toprule
			\multicolumn{1}{c}{}                                            
			& \multicolumn{4}{c}{\textbf{Memory Corruption Vulnerability}}
			\\
			\multicolumn{1}{l}{\textbf{CFI Variants}}
			& \multicolumn{1}{r}{\begin{tabular}[c]{@{}c@{}}\textsc{Heap-based}\\ \textsc{Buffer Overflow}\end{tabular}}
			& \multicolumn{1}{r}{\begin{tabular}[c]{@{}c@{}}\textsc{Stack-based}\\ \textsc{Buffer Overflow}\end{tabular}}
			& \multicolumn{1}{r}{\begin{tabular}[c]{@{}c@{}}\textsc{Use-After}\\ \textsc{-Free (UAF)}\end{tabular}}
			& \multicolumn{1}{r}{\begin{tabular}[c]{@{}c@{}}\textsc{Type}\\ \textsc{Confusion}\end{tabular}} 
			\\ 
			& \emph{CVE-2021-3156} 
			& \emph{CVE-2023-49992} 
			& \emph{CVE-2022-3666} 
			& \emph{CVE-2024-34391} 
            \\
            & \textsf{sudo}~\cite{sudo} 
			& \textsf{espreak-NG}~\cite{espeak-ng}
			& \textsf{Bento4}~\cite{bento4}
			& \textsf{libxmljs}~\cite{libxmljs}
			\\ \midrule
			
			\rowcolor{light-gray}
			\texttt{cfi-cast-strict}			&  
			$\circlearrowright$          	& 
			$\circlearrowright$          	& 
			$\circlearrowright$          	& 
			$\circlearrowright$          	\\ 
			
			\texttt{cfi-derived-cast}			& 
			$\circlearrowright$	            & 
			$\circlearrowright$             & 
			$\circlearrowright$             & 
			$\circlearrowright$             \\
			
			\rowcolor{light-gray}
			\texttt{cfi-unrelated-cast}		& 
			$\circlearrowright$          	& 
			$\circlearrowright$	            & 
			~${\circlearrowright}^{\textcolor{red}{\ast}}$	& 
			$\circlearrowright$             \\
			
			\texttt{cfi-nvcall}		& 
			$\circlearrowright$          	& 
			$\circlearrowright$          	& 
			$\circlearrowright$          	& 
			$\circlearrowright$          	\\
			
			\rowcolor{light-gray}
			\texttt{cfi-vcall}			& 
			$\circlearrowright$          	& 
			$\circlearrowright$             & 
			$\circlearrowright$          	& 
			$\circlearrowright$          	\\
			
			\texttt{cfi-icall}			&
			\textcolor{green}{\checkmark}	&
			\textcolor{green}{\checkmark}	&
			$\circlearrowright$            	&
			$\circlearrowright$          	\\
			
			\rowcolor{light-gray}
			\texttt{cfi-mfcall}	& 
			$\circlearrowright$          	& 
			$\circlearrowright$          	& 
			$\circlearrowright$          	& 
			$\circlearrowright$          	\\
			
			\bottomrule
		\end{tabular}
	\end{adjustbox}
    \begin{center}
            $\circlearrowright$ = compiling \textbf{but no} prevention of the specific CVE based PoC/exploit, 
            \textcolor{green}{\checkmark} = compiling \textbf{and} prevention\\
            ${}^{\textcolor{red}{\ast}}$\emph{compiling but CFI violation in code not connected to CVE (mismatch of old C/C++ standard~\cite{oldstandards1,oldstandards2})}
    \end{center}
    \vspace{-0.4cm}
\end{table*}
Table~\ref{tab:results} depicts the outcomes of our experiments based on the CVE collection process.

In a nutshell, the heap- and stack-based buffer overflow exploits were successfully mitigated by \texttt{icall}. 
While \texttt{unrelated-cast} prevented the UAF-PoC, the CFI violation occurred due to the library using an old C/C++ standard and not exploiting the vulnerability in itself.
CFI was not able to prevent the chosen type confusion exploit. 
The following explains the technical details of the chosen CVEs, the exploit process, and CFI's impact. 
We provide more in-depth technical details for each of the exploits in Appendix~\ref{sec:technical_details}.

\subsubsection{Heap-based Buffer Overflow - \emph{CVE-2021-3156}~\cite{CVE-2021-3156}}\label{sec:heap_exploit}
or ``\emph{Baron Samedit}'' is a heap overflow in \textsf{sudo}~\cite{sudo} allowing privilege escalation via malformed arguments ending with backslashes. The vulnerability bypasses \textsf{sudo}'s argument validation when using \texttt{sudoedit -s} instead of \texttt{sudo -e}~\cite{exploit-CVE-2021-3156}. 
The bug occurs in \texttt{set\_cmnd()} (Listing~\ref{lst:sudoers} in Appendix~\ref{sec:sudoedit}) where backslash handling increments pointers past buffer boundaries. 

\emph{\underline{CFI Impact:}} 
\texttt{register\_hooks()} causes a CFI violation of \textbf{\texttt{icall} preventing} the exploit in \texttt{load\_plugins.c}.
\texttt{register\_hooks()} is responsible for registering certain hooks involving function pointers needed to operate \textsf{sudo}. 
The heap buffer overflow caused by malformed input can overwrite these function pointers, effectively altering the execution flow.
When enabled, CFI ensures that indirect calls only jump to valid addresses explicitly registered or verified during the program's initialization. 
Since an attacker has altered the function pointer, CFI detects that the call is attempting to jump to an address that is not allowed.

\subsubsection{Stack-based Buffer Overflow - CVE-2023-49992~\cite{CVE-2023-49992}}\label{sec:stack_exploit}
describes a stack overflow in \textsf{eSpeak-NG}~\cite{espeak-ng}'s \texttt{RemoveEnding} function (Figure~\ref{fig:stack_buff} in Appendix~\ref{sec:espeak-ng}). 
A crafted input exploiting multibyte UTF-8 processing causes a buffer overflow in \texttt{ending[50]}, allowing out-of-bounds writes that corrupt stack variables and potentially hijack control flow, as demonstrated in the PoC for CVE-2023-49992.

\emph{\underline{CFI Impact:}} 
\textbf{\texttt{icall} prevents} this attack by enforcing strict runtime checks that verify whether an indirect function call actually points to a valid function of the expected type. 
A \texttt{Runtime Error in src/libespeak-ng/speech.c:473} in the function \texttt{synth\_callback} triggered by a CFI violation occurs.
When the corrupted \texttt{synth\_callback} function pointer is used at \texttt{speech.c}, the runtime detects that the pointer does not match a valid function of the expected type \texttt{int (short *, int, espeak\_EVENT *)}. 
Instead of allowing execution to proceed to an unintended location, \texttt{icall} immediately traps and aborts execution, preventing any malicious code from running. 
This stops the exploit before the attacker can gain control over the process. 
In summary, \texttt{icall} (i) shows a potential exploitation vector for the PoC and (ii) prevents it from being viable.

As predicted by our mapping in Table~\ref{tab:mapping}, the other CFI variants for function call integrity are less effective. 
\texttt{nvcall} protects non-virtual function calls, which are direct and unaffected by function pointer overwrites. 
\texttt{vcall} enforces integrity for virtual function calls in C++ objects, ensuring that calls to virtual functions only resolve to valid virtual table entries, which is irrelevant here since the exploit does not involve C++ virtual functions. 
\texttt{mfcall} protects member function calls on objects, ensuring that function pointers within objects are valid before invocation, but the function pointer, in this case, is not a member function but rather a standalone function pointer used in the callback mechanism.

\subsubsection{Use-After-Free - CVE-2022-3666~\cite{CVE-2022-3666}}\label{sec:uaf_exploit}
describes a UAF vulnerability in \textsf{Bento4}'s~\cite{bento4} \texttt{AP4\_LinearReader} class (still unpatched). 
When handling failed sample reads, the issue occurs in \texttt{Advance()}, see Listing~\ref{lst:ap4} in Appendix~\ref{sec:ap4}. 

\emph{\underline{CFI Impact:}} 
In \textsf{Bento4}, the vulnerability occurs when an object is deleted, but a dangling pointer to it remains accessible. 
If the same memory is later reallocated for a different object, an attacker can manipulate this stale pointer to interact with unintended memory contents. 
CFI primarily ensures that function calls and type casts follow the expected class hierarchy and valid control flow. 
Still, it does not prevent access to a freed object if the memory has been repurposed. 
In the available PoC, the stale pointer is not used to perform an exploit. 
Thus, the \texttt{vtable} is not corrupted, nor is an invalid cast performed. 
Moreover, the PoC leads to Denial-of-Service, which CFI does not cover. 
These are the reasons why \textbf{none} of the CFI variants detects a violation.

A CFI \texttt{unrelated-cast} violation occurs in \texttt{Ap4DArray.h}, which is not connected to CVE-2022-3666 but rather an incompatibility of CFI with older C/C++ standards~\cite{oldstandards1,oldstandards2}.

\subsubsection{Type Confusion - CVE-2024-34391~\cite{CVE-2024-34391}}\label{sec:tyconf_exploit}
describes a type confusion vulnerability in \textsf{libxmljs}~\cite{libxmljs}, the Node.js~\cite{nodejs} bindings for \texttt{libxml2}~\cite{libxml2}. 
The issue occurs during XML entity reference processing when SWIG-generated bindings incorrectly interpret an \texttt{xmlEntity} as an \texttt{xmlNode}, see Listings ~\ref{lst:xmlNode} and~\ref{lst:xmlEntity} in Appendix~\ref{sec:libxmljs}. 

\emph{\underline{CFI Impact:}} 
While all seven CFI variants compiled successfully, \textbf{none} of them was able to prevent the DoS exploit. 
The most likely variant to mitigate this kind of vulnerability would be one of the cast-based ones (see Table~\ref{tab:mapping}). 
These techniques require the presence of an actual cast, meaning a \texttt{dynamic\_cast} or \texttt{static\_cast} in C++. 
However, type confusion is not caused by an incorrect cast operation but rather by a misinterpretation of the type itself, more precisely, a misassumption of the memory layout of the accessed object. 
The problem arises when C function pointers are cast into C++ objects, circumventing proper type checks. 
Since the C++ compiler does not have full visibility on the memory layout and function pointer integrity of the underlying C code, it cannot enforce the expected control flow restrictions. 
This allows an attacker to exploit the type confusion and execute unintended behavior, bypassing CFI.

\section{Discussion}\label{sec:discussion}
\subsection{Effectiveness \& Limitations of CFI Preventing Exploitation}
In the cases where CFI prevents the exploitation of a specific vulnerability (Sections~\ref{sec:heap_exploit} and~\ref{sec:stack_exploit}), it may be possible to create a different exploit to circumvent CFI. 
However, we are not trying to show that CFI prevents exploitation altogether. 
Instead, our results show that in-the-wild exploits lose viability if CFI is used during compilation. 
In conjunction with other mitigation techniques, engineering an exploit to bypass all of them becomes increasingly complex. 
We also point out that in the cases of~\ref{sec:uaf_exploit} and~\ref{sec:tyconf_exploit}, even though CFI cannot stop these specific vulnerabilities, it does not mean CFI is unable to prevent exploitation of the underlying vulnerability type in general. 
Rather, our provided mapping (Table~\ref{tab:mapping}) depicts a general guideline. 

\subsection{Practical Guidance for Incremental CFI Deployment}
In practice, this means developers should not rely on a specific CFI variant to guarantee protection against exploitation of a vulnerability type. 
However, our mapping allows strategic planning of gradual CFI adoption to existing code bases. 
For example, one could start to integrate \texttt{vcall} and \texttt{icall} (as done by Chromium~\cite{chromiumcfi}), which already mitigates \textit{heap-based} and \textit{stack-based buffer overflow} exploits to a substantial degree while providing some protection against UAF. 

An alternate strategy is to focus on the CFI variants, which affect the usability of the target software the least. 
Let us assume compiling a piece of software with all CFI variants for type casting protection (Section \ref{sec:backgroundCFI}) works out of the box, while the other variants cause errors breaking the software's functionality due to compatibility issues, which is a common challenge of CFI~\cite{houy2024lessons,oldstandards1,oldstandards2,firefoxcfi}. 
Our mapping can help to find the best CFI option still offering the next most protection. 
Generally speaking, the more CFI variants are added, the better the protection. 
However, due to its highly complex implementation and compatibility challenges, it is recommended to start with the most useful ones, depending on the target project's vulnerabilities and needs.

\subsection{Challenges in Real-World CFI Application}
When applying CFI to real-world scenarios, unforeseen challenges may arise, as exemplified by the type confusion vulnerability discussed earlier in Section~\ref{sec:tyconf_exploit}. 
In this case, the issue originated from the interaction and, more specifically, the translation between multiple programming languages (C, C++, and JavaScript). 
Such challenges are less likely to manifest when working solely with synthetic or carefully constructed case-based examples, as these are designed to isolate specific factors rather than account for the complexities of real-world implementation. 
Existing literature has primarily focused on such controlled examples (see Section~\ref{sec:rw}), leaving gaps in understanding how these issues manifest in practical, real-world settings.

\subsection{Comparison with other CFI Implementations}
Windows Control Flow Guard (CFG)~\cite{windowscfi} provides a coarse-grained implementation of forward-edge CFI.
It restricts indirect calls to a set of valid function entry points, but does not enforce additional rules for signature matching or class hierarchies. 
GCC offers only limited CFI support compared to LLVM, as it focuses primarily on virtual table verification (VTV)~\cite{tice2014enforcing,vtv,gccSanitize,vtvProposal,vtvGuide}. 
This mechanism helps detect specific type confusion vulnerabilities in C++.
However, it does not cover other indirect control transfers, nor does it provide granularity of CFI variants. 
In contrast, LLVM implements a comprehensive CFI implementation that supports multiple variants, including checks on indirect calls, virtual functions, and various forms of type casting. 

Intel's Control Flow Enforcement Technology (CET)~\cite{CET} extends software CFI with hardware-based shadow stacks and indirect branch tracking, thereby strengthening protection against backward branches and improving security with minimal performance overhead. 
However, CET focuses primarily on backward-edge control flow integrity and does not replace the checks provided by software solutions such as LLVM's CFI. 
Given LLVM's flexible and detailed software-based options, it provides an ideal implementation for analyzing practical trade-offs in CFI deployment, which motivated our work on LLVM's CFI.

\section{Threats to Validity}\label{sec:valdility}
\subsection{Vulnerability Selection}~\label{sec:limit_collection}
We selected one representative CVE per vulnerability category from the KEV Top 10 list to ensure relevance and real-world impact.  
However, this small sample size may not capture the full extent of each category, as it is limited by the restricted availability of working PoCs and exploits. 
Thus, our conclusions may not be generalizable to all potential exploits in each class. 
Additionally, there may be other relevant memory corruption vulnerabilities (e.g., Time-of-check Time-of-use (TOCTOU) Race Condition) related to CFI that are less likely to occur in the real world and are therefore not included in the KEV list (and thus not in this work).

\subsection{Mapping from CFI Variants to Vulnerability Types}~\label{sec:limit_mapping}
Our taxonomy is based on the publicly available LLVM documentation, publications, and empirical testing. 
However, the literature is occasionally ambiguous or outdated, and not all implementation details are disclosed. 
Therefore, some mappings between CFI variants and vulnerability types required informed interpretation.

\subsection{Limitations of Experiment Setup}
Our evaluation used CVE PoCs and exploits executed on modified builds using CFI instrumentation. 
Although the only changes to the build options were related, we cannot guarantee that they did not have any other effects or that other potential modifications (e.g., using alternative optimization levels) could have an impact. 
Some exploits required minor adjustments to function in our testbed. Additionally, we did not include performance or usability evaluations, which are critical for real-world deployment but were outside the scope of this work.

\subsection{Human Intervention by the Researchers}
Significant manual effort was required to reproduce CVEs and determine whether CFI blocked exploitation. 
These judgments, although validated through debugging and reverse engineering, may still introduce human bias or error. 
In addition, we had to determine whether the identified exploits/PoCs fit our purpose, which could introduce an additional layer of bias, even though we made every effort to find representative and available options. 
Moreover, the mapping process, as described in Section~\ref{sec:limit_mapping}, required an informed interpretation of the source at specific points. 
We attempted to minimize bias by providing the codebook in the Appendix~\ref{sec:codebook}.

\section{Related Work}\label{sec:rw}
\subsection{CFI Deployment Challenges}
Several studies have identified key challenges in applying CFI in practice. 
Becker et al.~\cite{becker2024sok} highlight incomplete instrumentation, unprotected dynamic libraries, and partial enforcement that weaken security guarantees. 
Although this work is a SoK paper, it focuses solely on the actual deployment of CFI on Android and Windows rather than providing a guide on how and when to apply CFI. 
Houy and Bartel~\cite{houy2024lessons,houy2025twenty} emphasize performance overhead, compatibility with legacy code, and difficulties handling dynamic features like just-in-time compilation in complex runtimes. 
Xu et al.~\cite{xu2019confirm} point out frequent compatibility issues, false positives, and limited support for complex software patterns, revealing trade-offs between security precision and deployability. 
These challenges collectively hinder broad and effective CFI adoption.

\subsection{CFI Working Examples}
Nearly all of the papers rely on synthetic PoCs rather than real-world documented attacks. 
Evans et al.~\cite{evans2015control}, Niu et al.~\cite{niu2014modular,niu2015per}, Sayeed et al.~\cite{sayeed2019control}, and Xu et al.~\cite{xu2019confirm} evaluate LLVM's CFI using controlled experiments and synthetic exploit scenarios to demonstrate its effectiveness. 
Burow et al.~\cite{burow2017control} focus mainly on empirical testing with real-world applications and benchmark suites designed to simulate practical workloads and measure performance and security outcomes. 
In contrast, our work aims to provide a practical deployment guideline for developers, supporting informed and incremental adoption of CFI in existing code bases. 
Similarly, Conti et al.~\cite{conti2015losing}, and Li et al.~\cite{li2020finding} analyze CFI's impact on various attack vectors through constructed test cases. 
While Farkhani et al.~\cite{farkhani2018effectiveness} discuss attack techniques observed in practical settings, the evaluation still primarily relies on PoCs rather than exploits seen in the wild. 
These PoCs are carefully constructed experiments that illustrate how specific bypass techniques can be employed against LLVM's CFI, thereby exposing potential weaknesses in its implementation. 
Although the scenarios mimic real-world conditions to some extent, they remain controlled demonstrations rather than real-world attacks. 

\subsection{CFI Bypasses and Limitations}
Many attack vectors bypass coarse-grained CFI \cite{goktas2014out, davi2014stitching, carlini2014rop}. 
Carlini et al. showed that dispatcher functions (e.g., \texttt{memcpy}) can enable a \emph{return-to-libc} attack. 
However, CFI blocked arbitrary code execution in some cases \cite{carlini2015control}, with further reductions when combined with a shadow stack. 
ROP and JOP attacks use small code fragments for arbitrary execution. 
Studies~\cite{evans2015control,niu2014modular,niu2015per,burow2017control} indicate that LLVM's CFI limits naive ROP/JOP attacks by enforcing a fixed CFG, yet advanced memory disclosure bypasses remain. 
COOP attacks, which manipulate virtual function calls, also partially evade CFI protection~\cite{conti2015losing,schuster2015counterfeit,sayeed2019control}. 
Although enhanced per-input CFI improves forward-edge protection, LLVM's CFI still struggles against speculative execution-based attacks~\cite{xu2019confirm}. 
Similarly, shadow stacks strengthen defenses against stack-based and heap vulnerabilities~\cite{li2020finding,farkhani2018effectiveness} but do not entirely prevent return address manipulation.

\section{Conclusion}\label{sec:conclusion}
This work presents a structured guideline that maps the most critical memory corruption vulnerabilities to LLVM’s forward-edge CFI variants, offering developers a practical reference for informed CFI deployment. 
While CFI is a powerful mitigation, its adoption remains limited due to integration challenges, such as runtime violations caused by benign code patterns or mismatches with existing design decisions. 
Our guideline addresses these obstacles by clarifying which CFI variants are most suitable for which vulnerability types. 
To illustrate its relevance, we analyze four recent, high-impact vulnerabilities. 
In two cases, the mapped CFI variant would have blocked the exploitation, underscoring the guideline's practical utility.

We advocate for broader CFI adoption in memory-unsafe projects and call for future work to focus on easing its integration. Our guideline is a step toward making CFI more accessible and deployable in real-world systems.

\bibliographystyle{plain}
\bibliography{00_main}

\appendix
\subsection{Technical Details of Exploit Examples}\label{sec:technical_details}
\subsubsection{Heap-based Buffer Overflow - \emph{CVE-2021-3156}~\cite{CVE-2021-3156}}\label{sec:sudoedit}
Listing~\ref{lst:sudoers} shows the vulnerable code of \texttt{set\_cmd} function. 
\begin{listing}[hbt!]
\begin{lstlisting}[style=cppstyle]
static int set_cmnd(void) {
...
     for (to = user_args, av = NewArgv + 1; 
                (from = *av); av++) {
         while (*from) {
             if (from[0] == '\\' 
                    && !isspace((unsigned char)from[1]))
                 from++;
             *to++ = *from++;
         }
         *to++ = ' ';
     }
...
}
\end{lstlisting}
\caption{The vulnerable \texttt{set\_cmnd} function in \texttt{sudoers.c} (lines 10-11 show the faulty backslash handling)}
\label{lst:sudoers}
\end{listing}
When an argument ends with a backslash, \texttt{from[1]} accesses the null terminator, causing: (i) copying the null terminator and incrementing \texttt{from} out-of-bounds (line 11) and (ii) begins writing past the buffer end (line 13). 
Figure~\ref{fig:heap_buff} shows the exploit process. 
\begin{figure}[hbt!]
    \centering
    \includegraphics[width=.9\linewidth]{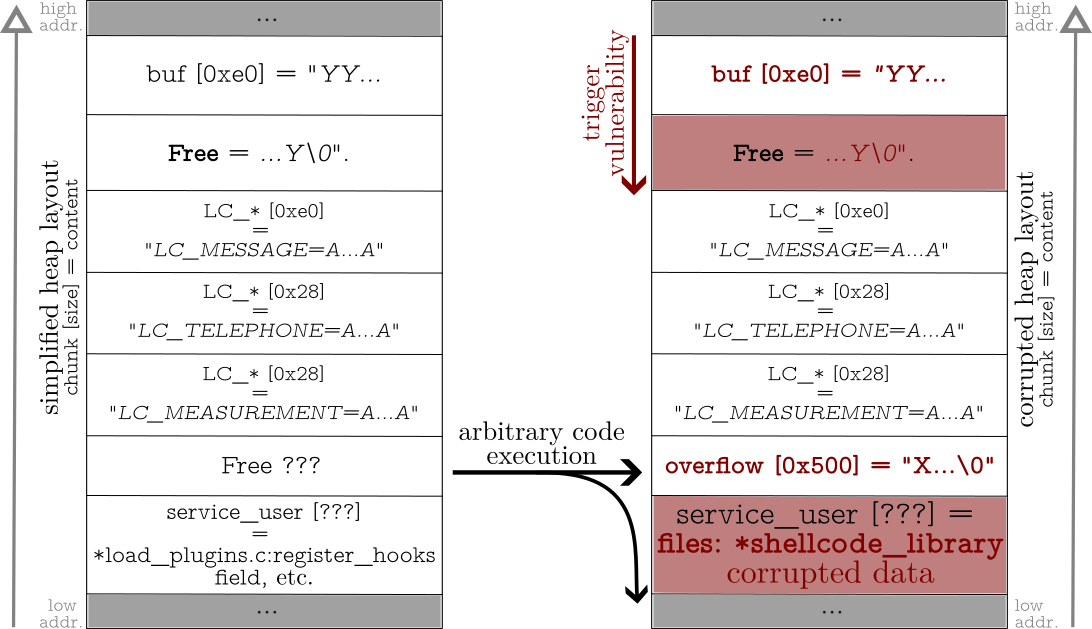}
    \caption{Heap layout during exploitation~\cite{exploit-CVE-2021-3156,sotirov2007heap}}
    \label{fig:heap_buff}
\end{figure}

The attack uses \texttt{LC\_*} variables for \emph{Heap Feng Shui}~\cite{sotirov2007heap}, overwriting a \texttt{service\_user} struct to load malicious library code and gain root access.

\subsubsection{Stack-based Buffer Overflow - CVE-2023-49992~\cite{CVE-2023-49992}}\label{sec:espeak-ng}
The vulnerability exploits UTF-8 multibyte handling to overflow a 50-byte \texttt{ending} buffer. 

The stack layout is critical (Figure~\ref{fig:stack_buff}): \texttt{ending[50]} sits below function parameters like \texttt{word\_copy} and the return address (\Circled[inner color=white, fill color=black]{1}). The overflow occurs when \texttt{len\_ending} (derived from \texttt{end\_type \& 0x3f}) exceeds 50. 
During multibyte UTF-8 processing (lead bytes: \texttt{0b11xxxxxx}, continuation: \texttt{0b10xxxxxx}~\cite{multibytes}), \texttt{word\_end} decrements out-of-bounds (\Circled[inner color=white, fill color=black]{2}), causing subsequent writes to corrupt adjacent stack variables. 
This allows overwriting \texttt{word\_copy} and potentially hijacking control flow. 
The PoC~\cite{exploit-CVE-2023-49992} demonstrates this via crafted input triggering a segmentation fault.
\begin{figure}[hbt!]
\centering
\includegraphics[width=\linewidth]{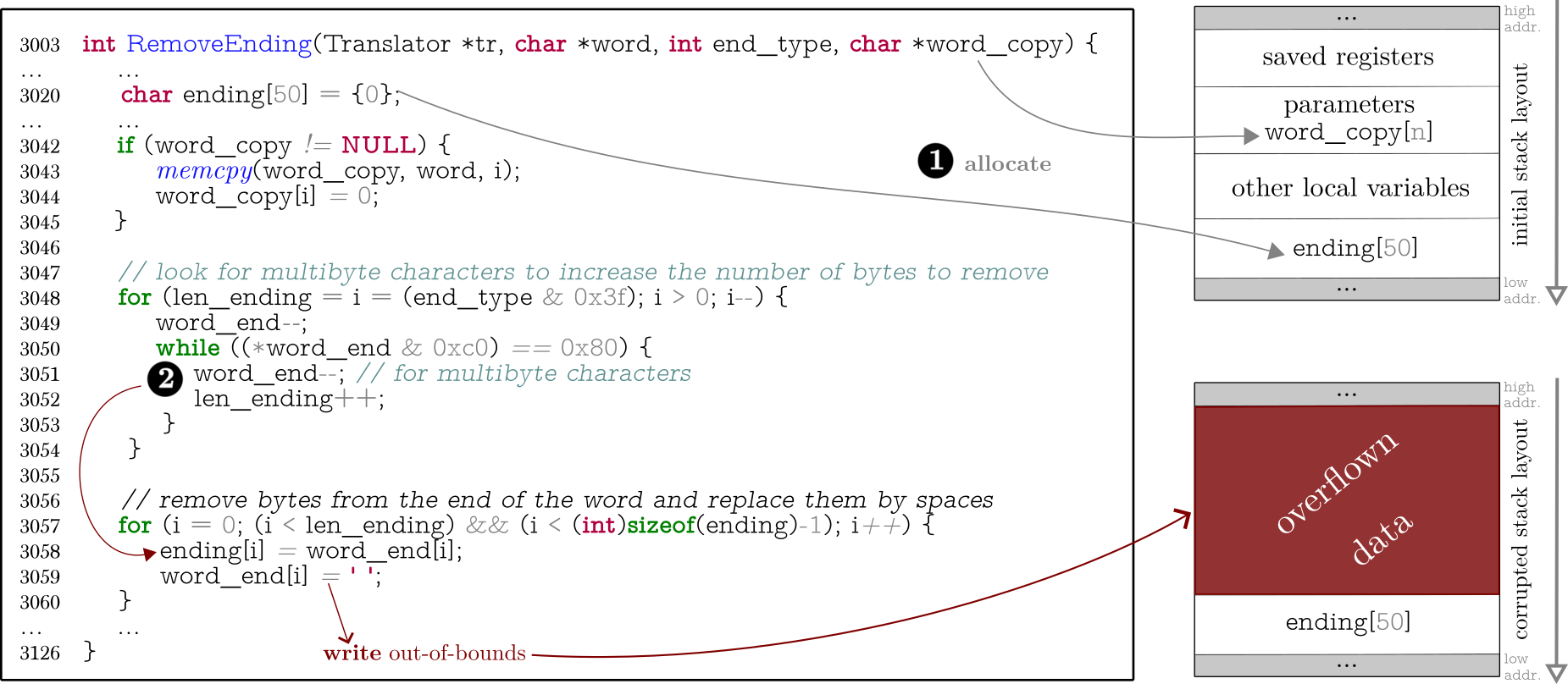}
\caption{Vulnerable \texttt{RemoveEnding} in \texttt{dictionary.c} (commit \texttt{b702b039})}
\label{fig:stack_buff}
\end{figure}

\subsubsection{Use-After-Free - CVE-2022-3666~\cite{CVE-2022-3666}}\label{sec:ap4}
The vulnerable code is displayed in Listing~\ref{lst:ap4}. 
The vulnerability manifests when (i) a sample allocation fails (line 8), freeing \texttt{m\_NextSample} but leaving the pointer unchanged, (ii) \texttt{SampleBuffer} takes ownership of \texttt{m\_NextSample} (line 13). 
(iii) A subsequent deletion of the buffer destroys the sample without nulling the original pointer, or (iv) the next \texttt{Advance()} call dereferences the dangling pointer. 
The PoC~\cite{exploit-CVE-2022-3666} triggers this via malformed MP4 files in \textsf{mp42ts}, exploiting the failed read path to cause memory corruption.
\begin{listing}[hbt!]
\begin{lstlisting}[style=cppstyle]
AP4_Result AP4_LinearReader::Advance(bool read_data) {
...
if (tracker->m_NextSample == NULL) {
 // Allocation
 tracker->m_NextSample = new AP4_Sample(); 
 // Freed but...
 if (AP4_FAILED(result)) 
    delete tracker->m_NextSample; 
}
// UAF
AP4_UI64 offset = tracker->m_NextSample->GetOffset();
...
SampleBuffer* buffer = new SampleBuffer(
                        next_tracker->m_NextSample);
if (AP4_FAILED(result)) {
 // Destroys m_NextSample via ~SampleBuffer()
 delete buffer; 
 // Dangling pointer remains
 return result; 
\end{lstlisting}
\caption{UAF vulnerability in \texttt{AP4\_LinearReader::Advance()}}
\label{lst:ap4}
\end{listing}

\subsubsection{Type Confusion - CVE-2024-34391~\cite{CVE-2024-34391}}\label{sec:libxmljs}
The vulnerability triggers when calling \texttt{attrs()} on an entity reference, causing the SWIG wrapper (\texttt{info.Holder()}) to incorrectly cast the object. 
This allows JavaScript to operate on the misidentified \texttt{xmlEntity} as if it were an \texttt{xmlNode}. 
The confusion stems from both structures sharing similar field layouts but differing at critical offsets - where \texttt{xmlNode} has a \texttt{properties} pointer, \texttt{xmlEntity} stores an integer \texttt{length} value (see Listings~\ref{lst:xmlNode} and~\ref{lst:xmlEntity}).

The impact depends on system configuration: On 32-bit systems with \texttt{XML\_PARSE\_HUGE} enabled, attackers may achieve RCE by manipulating memory layouts through oversized length values. 
In other cases (64-bit systems or without the flag), the vulnerability typically causes DoS via segmentation faults, as demonstrated in the available PoC~\cite{exploit-CVE-2024-34391}. 
The \texttt{XML\_PARSE\_HUGE} flag particularly exacerbates the issue by disabling size checks that would normally prevent such memory corruption.
\begin{center}
\begin{tabular}{m{0.45\columnwidth} m{0.45\columnwidth}}
\begin{minipage}{0.45\columnwidth}
\begin{lstlisting}[style=cppstyle]
struct _xmlNode {
  void *_private;
  xmlElementType type;
  ...
//Misinterpreted as length
  struct _xmlAttr 
               *properties;
  ...
};
\end{lstlisting}
\captionof{listing}{\texttt{xmlNode} structure}
\label{lst:xmlNode}
\end{minipage}
&
\begin{minipage}{0.45\columnwidth}
\begin{lstlisting}[style=cppstyle]
struct _xmlEntity {
  void *_private;
  xmlElementType type;
  ...
//Confused with properties
  int length;
  
  ...
};
\end{lstlisting}
\captionof{listing}{\texttt{xmlEntity} structure}
\label{lst:xmlEntity}
\end{minipage}
\end{tabular}
\end{center}

\subsection{Codebook of Mappings}\label{sec:codebook}
In Table~\ref{tab:codebook} the codebook used to create the mapping depicted in Table~\ref{tab:mapping} in Section~\ref{sec:mapping}. We provide the most important part of each reference as direct quotes. This does not mean that the rest of the paper is irrelevant. 

\onecolumn
\begin{center}
\tablefirsthead{%
\toprule
\textsc{Vulner-ability Type}& \textsc{Reference} & \textsc{Quotes}\\
\midrule}
\tablehead{%
\toprule
\textsc{Vulner-ability Type} & \textsc{Reference} &\textsc{Quotes} \\
\midrule}
\tabletail{%
\bottomrule
}
\tablelasttail{\bottomrule}
\tablecaption{CodeBook for Mapping}\label{tab:codebook}
\begin{supertabular}{P{2cm}|c|m{13cm}}

    \multirow{4}{*}{\parbox[c][0.5\textheight][c]{2cm}{\centering\rotatebox{90}{\textsc{H~~e~~a~~p~~-~~b~~a~~s~~e~~d~~~~~~B~~u~~f~~f~~e~~r~~~~~~O~~v~~e~~r~~f~~l~~o~~w}}}}

        &\cellcolor{light-gray}\footnotesize {\cite{abadi2009control}} &
        \cellcolor{light-gray}\begin{itemize}
            \item\footnotesize\emph{``CFI enforcement is effective against a wide range of common attacks, since abnormal control-flow modification is an essential step in many exploits—independently of whether buffer overflows and other vulnerabilities are being exploited [Pincus and Baker 2004; Wilander and Kamkar 2003]. We have examined many concrete attacks and found that CFI enforcement prevents most of them. These include both classic, stack-based buffer-overflow attacks and newer, heap-based “jump-to-\texttt{libc}” attacks.''}
            \item\footnotesize\emph{``On the other hand, CFI would have prevented all the other exploits that we studied because, in one way or another, they all endeavored to deviate from the expected control flow. Many exploits performed a “jump-to-\texttt{libc}” control transfer from a program point where this jump was not expected. Often this invalid control transfer was attempted through heap overflows or some form of pointer subterfuge (of the kind recently described by Pincus and Baker [2004]).''}
        \end{itemize}\\ 

        &\footnotesize{\cite{ammar2024bridging}} &
        \begin{itemize}
            \item\footnotesize\emph{``These attacks exploit memory corruption vulnerabilities, such as buffer overflows and use-after-free bugs, to achieve malicious objectives, exemplified by privilege escalation. Adversaries accomplish this by manipulating a program’s control flow through the execution of carefully selected code snippets, known as gadgets. ROP attacks are characterized by chaining a sequence of gadgets, each ending with a return (\texttt{ret}) instruction. On the other hand, JOP attacks construct chains of gadgets on forward edges by targeting code fragments that end with indirect \texttt{jump} or \texttt{call} instructions.}
            \newline\emph{\textbf{Problem Statement.} Over the past two decades, researchers have extensively studied Control Flow Integrity (CFI) schemes, which aim to mitigate control flow hijacking attacks by protecting both forward [13, 14, 17, 18, 35, 36] and backward [14, 18, 24, 36, 88] edges through enforcing various policies that must be respected during runtime.''}
            \item\footnotesize\emph{``These limitations arise from various factors, including the complex nature of target software [62], compatibility issues [60], and unbalanced trade-offs between security and performance [45]. Additionally, implementation mistakes represent a significant avenue for adversaries to exploit, bypassing detection by CFI defenses, which primarily operate at a local level [58, 62]. The concept of local detection within these mechanisms does not provide external evidence that could offer valuable insights into the execution flow of the target application, creating uncertainties regarding the current state of runtime integrity.''}
            \item\footnotesize\emph{``Considering the broader attack surface in complex software, early detection, similar to local CFI mechanisms, is preferable whenever feasible but without compromising the advantages of external evidence.''}
            \item\footnotesize\emph{``The few others that have been supported in mainstream compilers, e.g., LLVM CFI [17], are not comprehensive enough to completely prevent control flow violations [47, 52] or suffer from compatibility issues with some applications [60].''}
        \end{itemize}\\ 
        
        &\cellcolor{light-gray}\footnotesize{\cite{burow2017control}} & 
        \cellcolor{light-gray}\begin{itemize}
            \item\footnotesize\emph{``Memory corruptions, such as buffer overflows, are routinely exploited by attackers. Despite significant research into exploit mitigations, very few of these mitigations have entered practice [Szekeres et al. 2013]. The combination of three such defenses, (i) Address Space Layout Randomization (ASLR) [PaX-Team 2003a], (ii) stack canaries [van de Ven and Molnar 2004], and (iii) Data Execution Prevention (DEP) [Microsoft 2006], protects against \emph{code-injection attacks} but is unable to fully prevent \emph{code-reuse attacks}. Modern exploits use Return-Oriented Programming (ROP) or variants thereof to bypass currently deployed defenses and divert the control flow to a malicious payload. Common objectives of such payloads include arbitrary code execution, privilege escalation, and exfiltration of sensitive information. The goal of Control-Flow Integrity (CFI) [Abadi et al. 2005a] is to restrict the set of possible control-flow transfers to those that are strictly required for correct program execution. This prevents code-reuse techniques such as ROP from working because they would cause the program to execute control-flow transfers, which are illegal under CFI.''}
        \end{itemize}\\
                    
        &\footnotesize{\cite{conti2015losing}} & \begin{itemize}
            \item\footnotesize\emph{``For instance, both IFCC and VTV focus on heap-based attacks, like virtual-function table manipulation, assuming that the program stack and the return addresses it contains cannot be manipulated due to randomization and stack canaries [50]. However, this assumption is only valid for most stack-related vulnerabilities such as the classic stack-based buffer overflow [4]. We revisit this assumption and demonstrate new attacks against the stack using memory-corruption vulnerabilities that are unrelated to the stack.''}
            \item\footnotesize\emph{``Code-reuse attacks require the target program to contain memory corruption vulnerabilities such as buffer overflows that the adversary can exploit to hijack the control flow and redirect execution to existing code fragments already available in an application’s memory space. A prominent attack technique is return-oriented programming (ROP), which is based on maliciously combining and executing a chain of short code sequences of an application [44]. The key element is the return instruction which serves as connecting link for the various code sequences.}\newline
            \emph{One of the most promising defense techniques to thwart code-reuse attacks is control-flow integrity [1, 3].''} 
        \end{itemize}\\ 

    \multirow{6}{*}{\parbox[c][0.55\textheight][c]{2cm}{\centering\rotatebox{90}{\textsc{H~~e~~a~~p~~-~~b~~a~~s~~e~~d~~~~~~B~~u~~f~~f~~e~~r~~~~~~O~~v~~e~~r~~f~~l~~o~~w}}}}

        &\cellcolor{light-gray}\footnotesize{\cite{cowan1998stackguard}} & 
        \cellcolor{light-gray}\begin{itemize}
            \item\footnotesize\emph{``Notably, function pointers are highly susceptible to buffer overflow attack. An attacker could conceivably use a buffer overflow to overwrite a function pointer that is on the heap, pointing it to attack code injected into some other buffer on the heap. The attack code need not even overflow its buffer.''} 
        \end{itemize}\\

        &\footnotesize{\cite{goktas2014out}} &
        \begin{itemize}
            \item\footnotesize\emph{``Briefly, the exploit is based on a heap overflow that serves a dual purpose, so unlike other attacks, we only use a single vulnerability for the whole attack. First, it serves as a memory disclosure bug, which we use to locate the addresses of loaded components and of the gadgets we need to perform the attack. Second, it grants us control over a jump instruction. We redirect the controlled jump to a chain of gadgets linked through call instructions to corrupt the stack and, eventually, control a return instruction. We proceed to perform stack-pivoting [10] to be able to perform ROP using the gadgets available to us. Next, we use a chain of gadgets, linked through return instructions, to make a code segment writable and overwrite it with our shellcode. The final step executes one last, CFI-checked, but permissible, return instruction into our shellcode.''}
        \end{itemize}\\ 

        &\cellcolor{light-gray}\footnotesize{\cite{jia2017towards}} &
        \cellcolor{light-gray}\begin{itemize}
            \item\footnotesize\emph{``The allocation address is usually a heap address returned by memory management functions. However, the allocation size may be derived from user inputs. Even if it is not affected by inputs, e.g., developers use a constant number (e.g., 1020 at line 3) that seems to be big enough as the allocation size, the program may still be vulnerable to heap overflow.''}
        \end{itemize}\\ 

        &\footnotesize{\cite{heapbased}} &
        \begin{itemize}
            \item\footnotesize\emph{``Buffer overflows often can be used to execute arbitrary code, which is usually outside the scope of a program's implicit security policy. Besides important user data, heap-based overflows can be used to overwrite function pointers that may be living in memory, pointing it to the attacker's code. Even in applications that do not explicitly use function pointers, the run-time will usually leave many in memory. For example, object methods in C++ are generally implemented using function pointers. Even in C programs, there is often a global offset table used by the underlying runtime.''}
        \end{itemize}\\ 

        &\cellcolor{light-gray}\footnotesize{\cite{bufferoverflow}} &
        \cellcolor{light-gray}\begin{itemize}
            \item\footnotesize\emph{``The data sets the value of the return pointer so that when the function returns, it transfers control to malicious code contained in the attacker’s data. Although this type of stack buffer overflow is still common on some platforms and in some development communities, there are a variety of other types of buffer overflow, including Heap buffer overflow and Off-by-one Error among others.}
            \item\footnotesize\emph{\textbf{Platform}
            \begin{itemize}
                \item Languages: C, C++, Fortran, Assembly
                \item Operating platforms: All, although partial preventative measures may be deployed, depending on environment.''
            \end{itemize}}
        \end{itemize}\\ 
        
        &\footnotesize {\cite{sayeed2019control}} & 
        \begin{itemize}
            \item\footnotesize\emph{``A heap overflow fault is a type of buffer overflow where the buffer being overwritten is allocated in the heap portion of memory, generally meaning that the buffer was allocated using a routine such as \texttt{malloc()} [22,23]. [...]}\newline
            \emph{The only difference with respect to the stack-based overflow is the zone or area of memory where the overflow occurs. The heap is where long term global variables are allocated. While the stack is well defined and follows an invariant layout, the heap is a much more dynamic structure. The content of the heap depends on the data types used by the application (for example, an application written in C++ with a lot of virtual classes will have a heap full of function pointers), also the exact arrangement will depend on the DSA (Dynamic Storage Allocator [24]) that implements the \texttt{malloc()} service.''}
            \item\footnotesize\emph{``Indirect Function-Call Checks (IFCC) is a CFI transformation mechanism which is put into practice over LLVM 3.4 compiler. IFCC introduces a dynamic tool which can be used to analyze CFI and locate forward edge CFI vulnerabilities [43]. Its implementation policy primarily concentrates on three compiler-based mechanisms. First, it focuses on protecting data that is stored in the heap, second, it enforces forward edge CFI by enhancing two of the implemented mechanisms, and finally, it takes advantage of the run-time analysis tool which is designed to detect control-flow hijacking in the early stage of software development life cycle.''}
        \end{itemize}\\ 

    \multirow{3}{*}{\parbox[c][0.5\textheight][c]{2cm}{\centering\rotatebox{90}{\textsc{H~~e~~a~~p~~-~~b~~a~~s~~e~~d~~~~~~B~~u~~f~~f~~e~~r~~~~~~O~~v~~e~~r~~f~~l~~o~~w}}}}

        &\cellcolor{light-gray}\footnotesize {\cite{LLVMdoc}} &
        \cellcolor{light-gray}\footnotesize {EVERYTHING} \\

        &\footnotesize {\cite{tice2014enforcing}} &
        \begin{itemize}
            \item\footnotesize\emph{``This defense has been widely and successfully used to protect return addresses and stack data; hence, current attacks instead typically corrupt vtable and function pointers to subvert a forward edge (an indirect jump or call) in the control-flow graph. Forward edges can be protected using Control-Flow Integrity (CFI) but, to date, CFI implementations have been research prototypes, based on impractical assumptions or ad hoc, heuristic techniques.''}
            \item\footnotesize\emph{``Taking advantage of heap-based memory corruption bugs can allow an attacker to overwrite a function-pointer value, so that arbitrary machine code gets executed when that value is used in an indirect function call [6]. Such exploits are referred to as \emph{forward-edge} attacks, as they change forward edges in the program’s control-flow graph (CFG).''}
            \item\footnotesize\emph{``However, there has been a corresponding increase in attacks that corrupt program-control data stored on the heap, such as C++ vtable pointers (inside objects) [14], or function pointers embedded in data structures; these attacks subvert indirect control transfers and are known as “return-to-libc” or return-oriented programming [3, 8, 22, 25–27, 30].}\newline 
            \emph{In this paper we present three compiler-based mechanisms for further protecting the integrity of program control data. Focusing on the integrity of control-transfer data stored on the heap, two of our mechanisms enforce forward-edge CFI by restricting the permitted function pointer targets and vtables at indirect call sites to a set that the compiler, linker, and runtime have determined to be possibly valid. The third mechanism is a runtime analysis tool designed to catch CFI violations early in the software development life-cycle.''}
            \item\footnotesize\emph{``The vtables themselves are placed in read-only memory, so they cannot be easily attacked. However, the objects making the calls are allocated on the heap. An attacker can make use of existing errors in the program, such as use-after-free, to overwrite the vtable pointer in the object and make it point to a vtable created by the attacker. The next time a virtual call is made through the object, it uses the attacker’s vtable and executes the attacker’s code.''}
        \end{itemize}\\  
    
        &\cellcolor{light-gray}\footnotesize {\cite{zeng2015heaptherapy}} &
        \cellcolor{light-gray}\begin{itemize}
            \item\footnotesize\emph{``Therefore, runtime measures that protect program execution against overflow attacks are important. Such measures can be roughly divided into three categories: (1) Full execution monitoring; (2) Approaches that learn from history to improve themselves; and (3) Measures that greatly increase the difficulty of exploiting buffer overflows. Examples of the third category include StackGuard [15], Data Execution Prevention [2], Address Space Layout Randomization [49], [8], and concurrent heap scanning [52], [44].}\newline 
            \emph{The first category contains a variety of approaches, which range from bounds checking [21], [38], [6]; to shadow memory-based checking [25], [39], [5], [12]; to control/data flow monitoring [23], [4], [11]; to N-version/N-variant systems [16], [7], [30]. Even with many optimization methods they still lead to a very high overhead, or require significant extra computing resources.''} 
        \end{itemize}\\

\midrule

\multirow{1}{*}{\parbox[c][0.01\textheight][c]{2cm}{\centering\rotatebox{90}{\footnotesize\textsc{{Stack-based Buffer Overflow}}}}}

        &\footnotesize{\cite{abadi2009control}} &
        \begin{itemize}
            \item\footnotesize\emph{``CFI enforcement is effective against a wide range of common attacks, since abnormal control-flow modification is an essential step in many exploits—independently of whether buffer overflows and other vulnerabilities are being exploited [Pincus and Baker 2004; Wilander and Kamkar 2003]. We have examined many concrete attacks and found that CFI enforcement prevents most of them. These include both classic, stack-based buffer-overflow attacks and newer, heap-based “jump-to-\texttt{libc}” attacks.''} 
            \item\footnotesize\emph{``Figure 7 shows the original, vulnerable machine code relevant to the stack-based exploit. This exploit is enabled when a buffer overflow overwrites the comparison-function pointer \texttt{cmp} before it is passed to \texttt{qsort}. [...]}\newline 
            \emph{Figure 7 also shows the CFI instrumentation of \texttt{qsort}. As shown in the figure, the instrumentation includes IDs at return points and ID-checks before computed call instructions, but not before direct calls. (Although not shown in the figure, the instrumentation also includes IDs at the start of functions whose address is taken, such as the intended comparison function.) With CFI, the defect in the median function does not enable exploits. CFI forbids invalid control transfers into the middle of functions (as well as returns to the start of functions), and it, therefore, prevents the necessary first step of the exploits and would also prevent subsequent steps.''}
        \end{itemize}\\ 

    \multirow{4}{*}{\parbox[c][0.5\textheight][c]{2cm}{\centering\rotatebox{90}{\textsc{S~~t~~a~~c~~k~~-~~b~~a~~s~~e~~d~~~~~~B~~u~~f~~f~~e~~r~~~~~~O~~v~~e~~r~~f~~l~~o~~w}}}}

        &\cellcolor{light-gray}\footnotesize {\cite{ammar2024bridging}} &
        \cellcolor{light-gray}\begin{itemize}
            \item\footnotesize\emph{``These attacks exploit memory corruption vulnerabilities, such as buffer overflows and use-after-free bugs, to achieve malicious objectives, exemplified by privilege escalation. Adversaries accomplish this by manipulating a program’s control flow through the execution of carefully selected code snippets, known as gadgets. ROP attacks are characterized by chaining a sequence of gadgets, each ending with a return (\texttt{ret}) instruction. On the other hand, JOP attacks construct chains of gadgets on forward edges by targeting code fragments that end with indirect \texttt{jump} or \texttt{call} instructions. }\newline
            \emph{\textbf{Problem Statement.} Over the past two decades, researchers have extensively studied Control Flow Integrity (CFI) schemes, which aim to mitigate control flow hijacking attacks by protecting both forward [13, 14, 17, 18, 35, 36] and backward [14, 18, 24, 36, 88] edges through enforcing various policies that must be respected during runtime.''}
            \item\footnotesize\emph{``These limitations arise from various factors, including the complex nature of target software [62], compatibility issues [60], and unbalanced trade-offs between security and performance [45]. Additionally, implementation mistakes represent a significant avenue for adversaries to exploit, bypassing detection by CFI defenses, which primarily operate at a local level [58, 62]. The concept of local detection within these mechanisms does not provide external evidence that could offer valuable insights into the execution flow of the target application, creating uncertainties regarding the current state of runtime integrity.''}
            \item\footnotesize\emph{``Considering the broader attack surface in complex software, early detection, similar to local CFI mechanisms, is preferable whenever feasible but without compromising the advantages of external evidence.''}
            \item\footnotesize\emph{``The few others that have been supported in mainstream compilers, e.g., LLVM CFI [17], are not comprehensive enough to completely prevent control flow violations [47, 52] or suffer from compatibility issues with some applications [60].''} 
        \end{itemize}\\ 

        &\footnotesize{\cite{bierbaumer2018smashing}} &
        \begin{itemize}
            \item\footnotesize\emph{``There is a trend on maintaining the integrity of an application’s control flow at runtime. CFI is achieved by ensuring the integrity of forward and backward edges in the control flow graph. As we focus on stack based exploitation techniques targeting control flow information related to the backward edge, we only consider the backward edge validation relevant. Backward edge validation is typically done using a shadow stack [9].''} 
        \end{itemize}\\ 

        &\cellcolor{light-gray}\footnotesize{\cite{burow2017control}} &
        \cellcolor{light-gray}\begin{itemize}
            \item\footnotesize\emph{``By modifying return addresses on the stack, an attacker can “return” to all addresses in a program, the foundation of return-oriented programming [Shacham 2007; Checkoway et al. 2010; Roemer et al.2012].''}
            \item\footnotesize\emph{``Several attacks target data structures used by CFI mechanisms. StackDefiler [Contiet al. 2015] leverages the fact that many CFI mechanisms implement the enforcement as a compiler transformation. Due to this high-level implementation and the fact that the optimization infrastructure of the compiler is unaware of the security aspects, an optimization might choose to spill registers that hold sensitive CFI data to the stack, where it can be modified by an attack [Abadi et al. 2005b]. Any CFI mechanism will rely on some runtime data structures that are sometimes writeable (e.g., when MCFI loads new libraries and merges existing sets). Missing the Point [Evans et al. 2015a]shows that ASLR might not be enough to hide this secret data from an adversary.''}
        \end{itemize} \\
                    
        &\footnotesize{\cite{cowan1998stackguard}} &
        \begin{itemize}
            \item\footnotesize\emph{``Buffer overflow attacks exploit a lack of bounds checking on the size of input being stored in a buffer array. By writing data past the end of an allocated array, the attacker can make arbitrary changes to program state stored adjacent to the array. By far, the most common data structure to corrupt in this fashion is the stack, called a “stack smashing attack,” which we briefly describe here, and is described at length elsewhere [15, 17, 21].}\newline
            \emph{Many C programs have buffer overflow vulnerabilities, both because the C language lacks array bounds checking, and because the culture of C programmers encourages a performance-oriented style that avoids error checking where possible [14, 13]. For instance, many of the standard C library functions such as \texttt{gets} and \texttt{strcpy} do not do bounds checking by default.}\newline
            \emph{The common form of buffer overflow exploitation is to attack buffers allocated on the stack. Stack smashing attacks strive to achieve two mutually dependent goals, illustrated in Figure 1:}\newline
            \emph{\textbf{Inject Attack Code} The attacker provides an input string that is actually executable, binary code native to the machine being attacked. Typically this code is simple, and does something similar to \texttt{exec("sh")} to produce a \texttt{root} shell.} \newline
            \emph{\textbf{Change the Return Address} There is a stack frame for a currently active function above the buffer being attacked on the stack. The buffer overflow changes the return address to point to the attack code. When the function returns, instead of jumping back to where it was called from, it jumps to the attack code.''} 
        \end{itemize}\\ 

    \multirow{3}{*}{\parbox[c][0.65\textheight][c]{2cm}{\centering\rotatebox{90}{\textsc{S~~t~~a~~c~~k~~-~~b~~a~~s~~e~~d~~~~~~B~~u~~f~~f~~e~~r~~~~~~O~~v~~e~~r~~f~~l~~o~~w}}}}

        &\cellcolor{light-gray}\footnotesize{\cite{maunero2019cfi}} &
        \cellcolor{light-gray}\begin{itemize}
            \item\footnotesize\emph{``In [3], the enforcement of the \emph{Control-Flow Integrity} (CFI) as basic defense was formalized. CFI dictates that, during program execution, whenever a control-flow transfer occurs, it must target a valid destination, as determined by a \emph{Control-Flow Graph} (CFG) created at compile time. Several solutions for the CFI have been proposed [8] [22] [48] [35] [50] [30] [23] [42] [19] [21] [49]. Commonly, two phases are distinguished: during the \emph{offline} phase, the intended flow transfers of the program are computed, and in the \emph{online} phase it is verified whether these transfers are respected by the running program without divergencies. The offline phase is usually performed resorting to a static analysis of the program binary, finding its basic blocks. As in [32], a \emph{basic block} is defined as a linear sequence of program instructions having one entry (the first instruction executed) and no branches out except at the exit (a control-flow transfer instruction). All statements within a basic block are executed before transferring the control to the next basic block. Transfers and basic blocks assume the identity of edge s and vertices within the Control-Flow Graph, and if an online monitor (software or hardware) is able to guarantee that the program does not take paths different from those established in such a graph, then the program is considered secure and immune to redirection attacks.''} 
        \end{itemize}\\ 

        &\footnotesize{\cite{stackbased}} &
        \begin{itemize}
            \item\footnotesize\emph{``In a classic buffer overflow exploit, the attacker sends data to a program, which it stores in an undersized stack buffer. The result is that information on the call stack is overwritten, including the function’s return pointer. The data sets the value of the return pointer so that when the function returns, it transfers control to malicious code contained in the attacker’s data.''}
            \item\footnotesize\emph{``Although this type of stack buffer overflow is still common on some platforms and in some development communities, there are a variety of other types of buffer overflow, including Heap buffer overflow and Off-by-one Error among others. Another very similar class of flaws is known as Format string attack. There are a number of excellent books that provide detailed information on how buffer overflow attacks work, including Building Secure Software [1], Writing Secure Code [2], and The Shellcoder’s Handbook [3].''}
            \item\footnotesize\emph{``At the code level, buffer overflow vulnerabilities usually involve the violation of a programmer’s assumptions. Many memory manipulation functions in C and C++ do not perform bounds checking and can easily overwrite the allocated bounds of the buffers they operate upon. Even bounded functions, such as strncpy(), can cause vulnerabilities when used incorrectly. The combination of memory manipulation and mistaken assumptions about the size or makeup of a piece of data is the root cause of most buffer overflows.''}
            \item\footnotesize\emph{``\textbf{Platform}
            \begin{itemize}
                \item Languages: C, C++, Fortran, Assembly
                \item Operating platforms: All, although partial preventative measures may be deployed, depending on environment.''
            \end{itemize}} 
        \end{itemize}\\ 
        
        &\cellcolor{light-gray}\footnotesize {\cite{nicula2019exploiting}} &
        \cellcolor{light-gray}\begin{itemize}
            \item\footnotesize\emph{``The stack-based buffer overflow is one of the most commonly known vulnerabilities and it still one of the most exploited vulnerabilities that are affecting software and operating systems [1]. A successful exploitation of this vulnerability can sometimes be difficult to achieve, and modern operating system nowadays have protection mechanisms in-place in order to prevent such issues to be exploited. These protections can also be implemented at the binary level in order to increase its security level. However, certain techniques can be used in order to bypass these prevention mechanisms, but all the techniques described do need auxiliary information in order to be implemented. The study presented will be focused on Intel architecture x86-x64, being more targeted around the Linux operating system internals and having as the main scope achieving code execution on the underlying operating system.''}
            \item\footnotesize\emph{``A good example would be a buffer overflow vulnerability that can be exploited in order to point the instruction pointer to a specific address inside the stack which is controlled by our buffer input. In this scenario, even though we have control on the Instruction Pointer, we cannot execute data that is being held in the stack frame which we overwrite. In order to bypass this protection, a technique called Return Oriented Programming (or ROP chain) can be used. By using this technique, which can also be reference as Return to libc attack [8], we can bypass the DEP protection [9] by re-using code already present in the exploited binary. Sometimes, the studied binary does not have all the needed instructions inside its base-code in order to fully exploit an existing buffer overflow. As such, the libc attack can be used. Inside the libc library, we can re-use a variety of instructions to fulfil the scope of exploiting the vulnerability. To use the libc code-base, we need to further leak an address inside the targeted binary. This can be achieved, for example, by chaining a format string vulnerability affecting the vulnerable binary.''}
            \item\footnotesize\emph{``A stack-based buffer overflow can be exploited in multiple ways depending on several variables. First, the allocated buffer size can play a huge role in choosing the right way to exploit the issue. In the previous case study shown, the buffer size was generous and allowed us enough space to inject various addresses and use multiple techniques without worrying too much about the memory space. If the buffer size was restricted to a small limited number of characters, additional steps would be required to successfully exploit the vulnerability. For example, an additional input buffer may have been required to redirect the execution flow into it however, that hypothetical input needed to be, again, controlled by the adversary. Some techniques used to get around the limited buffer size promote the usage of environmental variables that are loaded by the binary when executed. A memory leak address is needed in order to obtain such details. The second problem raised is related to the protection mechanisms that are preventing a straight forward exploitation technique.''}
        \end{itemize}\\ 

    \multirow{6}{*}{\parbox[c][0.75\textheight][c]{2cm}{\centering\rotatebox{90}{\textsc{S~~t~~a~~c~~k~~-~~b~~a~~s~~e~~d~~~~~~B~~u~~f~~f~~e~~r~~~~~~O~~v~~e~~r~~f~~l~~o~~w}}}}

         &\footnotesize{\cite{bufferoverflow}} &
        \begin{itemize}
            \item\footnotesize\emph{``Buffer overflows often can be used to execute arbitrary code, which is usually outside the scope of a program's implicit security policy.''} 
        \end{itemize}\\ 
        
        &\cellcolor{light-gray}\footnotesize{\cite{padaryan2015automated}} &
        \cellcolor{light-gray}\begin{itemize}
            \item\footnotesize\emph{``If the size of data written to the buffer exceeds the buffer size, then the memory above the buffer on the stack will be rewritten; in particular, the old ebp value, the return address, and the function arguments can be rewritten. If the attacker can control the values written to the buffer, then he or she can ensure that the return address will contain the pointer to an arbitrary code, which can be formed by the attacker. The execution of such a code can have serious repercussions, including the compromise of the application data, the user, or the operating system.''}
            \item\footnotesize\emph{``However, often one of the registers points to a stack space at the time of returning from the function. If this space is available for code placing, then the control can be transferred to this code using the instruction \texttt{jmp 〈reg〉} or \texttt{call 〈reg〉} that is located at a known address. The instructions of this type are called trampolines. In this case, upon the return from the function the control will be transferred to the trampoline instruction, and from it to the code placed on the stack (Fig. 1c).''}
        \end{itemize} \\ 

        &\footnotesize{\cite{sayeed2019control}} &
        \begin{itemize}
            \item\footnotesize\emph{``A stack-based buffer overflow fault is a type of out-of-bounds write where the buffer being overwritten is allocated on the stack (i.e., a local variable or, rarely, a parameter to a function) [21]. [...] As we will see later, the exploitability of a fault greatly depends on the type of data that can be overwritten; stack buffer overflow is one of the simplest to exploit due to the presence of the return address.''}
            \item\footnotesize\emph{``For this reason, the stack was, and still is, both very appealing to attackers and security researchers to develop protection techniques. Listing 8 shows the end of a function with a return (\texttt{retq}) at the end. A complete example of this exploitation was presented at the Black Hat conference in 2018 [8].''}
        \end{itemize}\\ 

        &\cellcolor{light-gray}\footnotesize{\cite{LLVMdoc}}
        &\cellcolor{light-gray}\footnotesize{EVERYTHING} \\

        &\footnotesize{\cite{tice2014enforcing}} &
        \begin{itemize}
            \item\footnotesize\emph{``This is perfectly correct behavior since the first function call is guaranteed by the platform ABI to preserve the value of the register. However, if \texttt{rbx} is spilled to the stack in \texttt{foo()} and is later overwritten, e.g., via a buffer overflow on the stack, then the call to \texttt{bar()} will be to an attacker-controlled location. An alternative to using a callee-saved register is to explicitly spill/reload the register to/from the stack, which has similar security concerns.''} 
        \end{itemize}\\ 
        
        &\cellcolor{light-gray}\footnotesize{\cite{xu2022bofaeg}} &
        \cellcolor{light-gray}\begin{itemize}
            \item\footnotesize\emph{``However, ROP (Return-oriented Programming) [5] attack can effectively bypass NX. It is a new type of attack based on code reuse technology, in which attackers extract instruction fragments (gadgets) from existing libraries or executable files to build malicious code. These instructions themselves are located in executable text segments, and they end with ret instructions to realize the convergence of instruction segment execution flows.}
            \item\footnotesize\emph{Libc.so is a dynamically linked version of the runtime glibc in the C language library under Linux. And it contains a large number of functions that can be used, including system, execve. Therefore, the attacker can gain control of the target program by finding the addresses of these functions in memory and using the stack buffer vulnerability to overwrite the return addresse to these functions.''}
            \item\footnotesize\emph{``The address of the libc.so function introduced by the program is stored in the GOT table, and the PLT table transfers the program control flow to the actual function by referencing the function address in the GOT table. The right shows that after leaking the address, it places the “/bin/sh” string and the address of system in the stack.''}
            \item\footnotesize\emph{``It first leaks the address of puts in libc.so by using the puts function introduced in the program, so that the address of libc.so loaded in the program can be obtained. *en it transfers control flow back to vuln\_func to trigger the stack buffer overflow vulnerability again. Finally, after getting the load address of libc.so, it sets the function parameter to the address of “/bin/sh”, and jumps to the system for execution, where “/bin/sh” and system are located in libc.so. That is, the program control flow is hijacked from the program code segment to libc.so for execution.''}
            \item\footnotesize\emph{``Therefore, Linux adopts the delayed binding mechanism, the basic idea of which is that the function is bound (symbol search, relocation, etc.) when it is first used, and it is not bound if it is not used. The address in the GOT table is obtained by relocation through \_dl\_runtime\_resolve when the function is used for the first time. Therefore, the main idea of ret to dl-resolve is to forge the relocation structure of the target function (including function name, relocation table) in the readable and writable memory. Then it uses \_dl\_runtime\_resolve to resolve the target function (such as system) and hijack the program control flow to the target function.''}
        \end{itemize}\\ 

        \multirow{1}{*}{\parbox[c][0.01\textheight][c]{2cm}{\centering\rotatebox{90}{\begin{tabular}[c]{@{}c@{}}\footnotesize\textsc{Stack-based}\\\footnotesize\textsc{Buffer Overflow}\end{tabular}}}}
        
        &\footnotesize {\cite{zhang2013practical}} &
        \begin{itemize}
            \item\footnotesize\emph{``Direct \emph{jmp/call} and conditional jump (\emph{jo}, \emph{jz}, etc.) instructions. Most \emph{jmp/call} instructions fall into this category. Their targets are fixed in the code, so DEP or W$\oplus$X protection prevents attackers from tampering with them.
            \item Indirect \emph{jmp/call} instructions, as used for function pointers and virtual method dispatch. Their targets usually are read from memory and may be controlled by attackers.
            \item All \emph{ret} instructions. Their targets are computed and pushed onto the stack at run-time by corresponding call instructions. Attackers may overwrite the return address on the stack to launch attacks like ROP and return-to-libc. The first two kinds of transfers are already well protected, [...].''}
        \end{itemize}\\

\midrule

    \multirow{4}{*}{\parbox[c][0.65\textheight][c]{2cm}{\centering\rotatebox{90}{\textsc{U~~s~~e~~-~~A~~f~~t~~e~~r~~-~~F~~r~~e~~e~~~~~~(~~U~~A~~F~~)}}}}
    
        &\cellcolor{light-gray}\footnotesize{\cite{abadi2009control}} &
        \cellcolor{light-gray}\begin{itemize}
            \item\footnotesize\emph{``For each thread of execution, this initial code can truncate the existing code and data segments and specify that the new, isolated segment lies within the freed-up address region. CFI can guarantee that the machine code for this set-up activity will remain inaccessible once it has executed.''} 
        \end{itemize}\\
        
        &\footnotesize{\cite{ammar2024bridging}} &
        \begin{itemize}
            \item\footnotesize\emph{``These attacks exploit memory corruption vulnerabilities, such as buffer overflows and use-after-free bugs, to achieve malicious objectives, exemplified by privilege escalation. Adversaries accomplish this by manipulating a program’s control flow through the execution of carefully selected code snippets, known as gadgets. ROP attacks are characterized by chaining a sequence of gadgets, each ending with a return (\texttt{ret}) instruction. On the other hand, JOP attacks construct chains of gadgets on forward edges by targeting code fragments that end with indirect \texttt{jump} or \texttt{call} instructions.}\newline 
            \emph{\textbf{Problem Statement.} Over the past two decades, researchers have extensively studied Control Flow Integrity (CFI) schemes, which aim to mitigate control flow hijacking attacks by protecting both forward [13, 14, 17, 18, 35, 36] and backward [14, 18, 24, 36, 88] edges through enforcing various policies that must be respected during runtime.''}
            \item\footnotesize\emph{``These limitations arise from various factors, including the complex nature of target software [62], compatibility issues [60], and unbalanced trade-offs between security and performance [45]. Additionally, implementation mistakes represent a significant avenue for adversaries to exploit, bypassing detection by CFI defenses, which primarily operate at a local level [58, 62]. The concept of local detection within these mechanisms does not provide external evidence that could offer valuable insights into the execution flow of the target application, creating uncertainties regarding the current state of runtime integrity.''}
            \item\footnotesize\emph{``Considering the broader attack surface in complex software, early detection, similar to local CFI mechanisms, is preferable whenever feasible but without compromising the advantages of external evidence.''}
            \item\footnotesize\emph{``The few others that have been supported in mainstream compilers, e.g., LLVM CFI [17], are not comprehensive enough to completely prevent control flow violations [47, 52] or suffer from compatibility issues with some applications [60].''} 
        \end{itemize}\\ 
        
        &\cellcolor{light-gray}\footnotesize{\cite{conti2015losing}} &
        \cellcolor{light-gray}\begin{itemize}
            \item\footnotesize\emph{``\textbf{Memory read-write:} The target program contains a memory-corruption vulnerability that allows the adversary to launch a run-time exploit. In particular, we focus on vulnerabilities that allow the adversary to read (information disclosure) and write arbitrary memory. Such vulnerabilities are highly likely as new vulnerabilities are being constantly reported. Common examples are use-after-free errors [49].''}
        \end{itemize}\\ 

        &\footnotesize {\cite{jia2017towards}} &
        \begin{itemize}
            \item\footnotesize\emph{``Although this off-by-one vulnerability could only overwrite one extra byte of 0, it is still exploitable. For example, in the running example, it will cause a further use-after-free vulnerability and lead to control flow hijacking. Details are omitted due to the space limitation.''} 
        \end{itemize}\\ 
        
        &\cellcolor{light-gray}\footnotesize{\cite{lee2015preventing}} &
        \cellcolor{light-gray}\begin{itemize}
            \item\footnotesize\emph{``In addition, Control Flow Integrity (CFI) tools can be used to prevent use-after-free vulnerabilities from being exploited to hijack the control-flow because the majority of vulnerability exploitations hijack the control flow to execute malicious code with Return-Oriented Programming (ROP) [35]. However, due to the inherent limitations of these tools, most of them only enforce coarse-grained CFI, which leaves some control-flows exploitable [7, 11 , 15 , 16]. Moreover, since control-flow hijacks are not the only method to compromise a program, it is still possible to bypass these techniques even if they can enforce perfect CFI, e.g., via non-control data attacks [8, 27].''}
            \item\footnotesize\emph{``Control flow integrity (CFI) [1 , 49 –51] enforces indirect function calls to be legitimate (i.e., enforcing integrity of the control-flows). Similarly, SafeDispatch [22]prevents illegal control flows from virtual function call sites. Unlike use-after-free and memory error detectors, CFI makes use-after-free vulnerabilities difficult to exploit. Specifically, CFI only shepherds function pointers to guarantee legitimate control flows. In practice, however, most CFI implementations enforce coarse-grained CFI to avoid heavy performance overheads and false positive alarms, but recent research [7, 11, 15, 16] has demonstrated that all the aforementioned coarse-grained CFI implementations can be bypassed. Moreover, dangling pointers can also be abused to corrupt non-control data (e.g., vector length variables, user privilege bits, or sandbox enforcing flags) in objects [8], all of which are not function pointers, which makes CFI based protection techniques bypassable. For example, a recent attack [27] overwrote user permission bits in the metadata to bypass user authorizations, including all other defense mechanisms. As an another example, vector length variable corruption is one popular technique to exploit use-after-free vulnerabilities that lead to information leakage attacks or additional heap buffer overflows.''} 
        \end{itemize}\\ 

    \multirow{6}{*}{\parbox[c][0.5\textheight][c]{2cm}{\centering\rotatebox{90}{\textsc{U~~s~~e~~-~~A~~f~~t~~e~~r~~-~~F~~r~~e~~e~~~~~~(~~U~~A~~F~~)}}}}

        &\footnotesize {\cite{li2020finding}} &
        \begin{itemize}
            \item\footnotesize\emph{``\emph{Use after free (UAF)}. This is a typical type of temporal vulnerabilities, in which a dangling pointer refers to a memory region allocated to a new object. Dereferencing the dangling pointer will corrupt the new object.''}
            \item\footnotesize\emph{``\textbf{Table 2: Effectiveness of CFI mechanisms against typical attack scenarios. The left-most column represents the specific attack scenario, and the right-most column represents the detail type of attack. The meaning of each symbol is shown in the label below the table.}''}
            \item\footnotesize\emph{``If the member variable is initialized in the base class, then in the VTable reuse attack that exploits the UAF vulnerability, the hijacked ICT only allows jumping to the base class instead of the same class.''}
        \end{itemize}\\ 

        &\cellcolor{light-gray}\footnotesize {\cite{uaf}} &
        \cellcolor{light-gray}\begin{itemize}
            \item\footnotesize\emph{``The use of previously freed memory may corrupt valid data, if the memory area in question has been allocated and used properly elsewhere.''}
            \item\footnotesize\emph{``If malicious data is entered before chunk consolidation can take place, it may be possible to take advantage of a write-what-where primitive to execute arbitrary code. If the newly allocated data happens to hold a class, in C++ for example, various function pointers may be scattered within the heap data. If one of these function pointers is overwritten with an address to valid shellcode, execution of arbitrary code can be achieved.''}  
        \end{itemize}\\ 

        &\footnotesize{\cite{owasp-uaf}} &
        \begin{itemize}
            \item\footnotesize\emph{``The use of heap allocated memory after it has been freed or deleted leads to undefined system behavior and, in many cases, to a write-what-where condition.''}
            \item\footnotesize\emph{``Access Control (instruction processing): If malicious data is entered before chunk consolidation can take place, it may be possible to take advantage of a write-what-where primitive to execute arbitrary code.''}
            \item\footnotesize\emph{``\textbf{Platform} 
            \begin{itemize}
                \item Languages: C, C++, Assembly
                \item Operating Platforms: All''
            \end{itemize}}  
        \end{itemize}\\ 
        
        &\cellcolor{light-gray}\footnotesize{\cite{sayeed2019control}} &
        \cellcolor{light-gray}\begin{itemize}
            \item\footnotesize\emph{`This is also referred to as a dangling pointer [31].  The process uses dynamic memory after it was freed.  In order to manifest the error, the program must free a valid object, then allocate a new one (it does not need to be of the same type) and then use the former pointer.  Although it may seem an unlikely scenario, there are two common situations: Error conditions and other exceptional circumstances that free memory but still perform some actions with the data or confusion over which part of the program is responsible for freeing the memory. This memory vulnerability is listed on the Common weakness Enumeration list as CWE-416, [...] dynamic memory is allocated in the heap, in this sense it can be considered to be a form of heap buffer overflow with additional limitations due to the amount of data that can be written(only the size of the new allocated object).''} 
        \end{itemize}\\  

        &\footnotesize{\cite{LLVMdoc}} &
        \footnotesize{EVERYTHING} \\ 
        
        &\cellcolor{light-gray}\footnotesize{\cite{tice2014enforcing}} &
        \cellcolor{light-gray}\begin{itemize}
            \item\footnotesize\emph{``To make these attacks more concrete, consider a C++ program that makes virtual calls and has a use-after-free bug involving some object. After the object is freed, an attacker can reallocate the memory formerly occupied by the object, overwriting its vtable pointer. Later virtual calls through this object get the attacker’s vtable pointer and jump to a function from the attacker’s vtable. Such exploits are becoming commonplace, especially for web browsers where the attacker can partially control executed JavaScript code [14, 23].}\newline
            \emph{Control-Flow Integrity (CFI) [1] guards against these control-flow attacks by verifying that indirect control-flow instructions target only functions in the program’s CFG.''}
            \item\footnotesize\emph{``However, the objects making the calls are allocated on the heap. An attacker can make use of existing errors in the program, such as use-after-free, to overwrite the vtable pointer in the object and make it point to a vtable created by the attacker. The next time a virtual call is made through the object, it uses the attacker’s vtable and executes the attacker’s code.''} 
        \end{itemize}\\ 

        \multirow{1}{*}{\parbox[c][0.01\textheight][c]{2cm}{\centering\rotatebox{90}{\begin{tabular}[c]{@{}c@{}}\footnotesize\textsc{Use-After-Free}\\\footnotesize\textsc{(UAF)}\end{tabular}}}}
        
        &\footnotesize {\cite{zeng2015heaptherapy}} &
        \begin{itemize}
            \item\footnotesize\emph{``While the focus of this paper is the heap buffer overflow problem, HeapTherapy can generate patches to deal with many other memory errors, such as double frees, dangling pointers and uninitialized heap buffer read. We can extend the specification of the temporary patch to support more bug types, and add bug-specific predefined handling to the memory allocation wrapper. We discuss some examples in the section. One type of dangling pointer bugs is due to a premature deallocation of some buffers. Assume the temporary patch that fixes a dangling pointer bug is generated. We can code a simple handling, which is invoked when a \texttt{free} call is hooked: the handling identifies whether the buffer’s CCID matches the VCCID in the patch; if so, the buffer’s deallocation is delayed.''}  
        \end{itemize}\\

\midrule

    \multirow{1}{*}{\parbox[c][0.03\textheight][c]{2cm}{\centering\rotatebox{90}{\textsc{T~~y~~p~~e~~~~~~C~~o~~n~~f~~u~~s~~i~~o~~n}}}}

        &\cellcolor{light-gray}\footnotesize{\cite{badoux2025type++}} &
        \cellcolor{light-gray}\begin{itemize}
            \item\footnotesize\emph{``With polymorphic objects (those with virtual functions), C++ introduces vtable pointers (\emph{vptr}), \emph{i.e.}, runtime type information (RTTI) to uniquely identify the type of the object. Without vtable pointer, the type of a memory area remains opaque at runtime. C++ provides different cast operations offering varying guarantees. Only dynamic casts execute a runtime check, but can only be used for polymorphic objects since they rely on the RTTI of vtable pointers. Conversely, static casts leverage the declared types in the source code to statically check class relationships while C-style casts carry neither a dynamic nor static check. To enforce type safety, only \texttt{dynamic\_cast} should be used [4], an option that core contributors of the standard envision to ease even at the cost of breaking the ABI [5].}\newline
            \emph{LLVM-CFI [6] leverages RTTI to protect casts involving the fraction of polymorphic objects, \emph{i.e.}, only around 3\% of all the casts in our benchmarks.''}
            \item\footnotesize\emph{``Natively, LLVM-CFI performs type checks only over polymorphic objects, thus leaving a wide attack surface for all non-polymorphic object types.''}
            \item\footnotesize\emph{``Similarly to LLVM-CFI, type++ has an option to additionally protect unrelated casts (\emph{i.e.}, cast over \texttt{void*}) whose classes are polymorphic or instrumented.''}
            \item\footnotesize\emph{``For instance, OMNeT++ 2006 shows a performance overhead of 4.07\% while protecting around 800M more casts than LLVM-CFI when considering \emph{down-to-cast} and \emph{unrelated casts} together (around 2.5B casts in total for type++ against 1.9B casts for LLVM-CFI, i.e., 30\% more).''}
            \item\footnotesize\emph{``Chromium contains 3,201 classes involved in down-casting, among which 2,099 are in necessary for type++ to achieve Property 2. LLVM-CFI already protects 1,102 classes (34.4\%).''}
            \item\footnotesize\emph{``Our evaluation uses the average of 10 runs of JetStream2 [28], testing three Chromium configurations: baseline, LLVM-CFI, and type++ with support of 1,928 classes (94.6\% of the total). Table VI shows the final results. JetStream2 reported a score of 96.14 for baseline Chromium while LLVM-CFI suffers a 0.43\% score reduction (95.73).''}
            \item\footnotesize\emph{``When evaluating LLVM-CFI on Chromium, we observe it detects only 347M (41.8\% of the total) down-to casts.''}
            \item\footnotesize\emph{``Using LLVM-CFI cast protection, all the polymorphic classes become protected, \emph{i.e.}, 34.4\% of the total.''}
            \item\footnotesize\emph{``For instance, LLVM-CFI [6] and UBSan [43] perform type checks at runtime for polymorphic objects involved in unsafe casts. However, both tools only deal with polymorphic classes while ignoring Plain Old Data objects.''}  
        \end{itemize}\\ 

    \multirow{2}{*}{\parbox[c][0.4\textheight][c]{2cm}{\centering\rotatebox{90}{\textsc{T~~y~~p~~e~~~~~~C~~o~~n~~f~~u~~s~~i~~o~~n}}}}

        &\footnotesize{\cite{cowan1998stackguard}} &
        \begin{itemize}
            \item\footnotesize\emph{``All of the vulnerabilities described here result from the lack of type safety in C. If the only operations that can be performed on a variable are those described by the type, then it is not possible to use creative input applied to variable \texttt{foo} to make arbitrary changes to the variable \texttt{bar}.''}  
        \end{itemize}\\

        &\cellcolor{light-gray}\footnotesize{\cite{kim2024bintyper}} &
        \cellcolor{light-gray}\begin{itemize}
            \item\footnotesize\emph{``The C++ language supports typecast operations to convert the type of variables. There are four types of typecast operations: \emph{reinterpret\_cast}, \emph{static\_cast}, \emph{dynamic\_cast}, and \emph{C-style cast}. \emph{reinterpret\_cast} is used in the form of \emph{reinterpret\_cast<destination\_type>} \emph{(source\_variable)}. This typecast operation does not verify whether conversion between \emph{source\_variable} type and \emph{destination\_type} is possible (typecasting compatibility). \emph{static\_cast} is used in the form of \emph{static\_cast<destination\_type> (source\_variable)}. \emph{static\_cast} verifies \emph{source\_variable} type and \emph{destination\_type} compatibility at compile time. The \emph{static\_cast} checks whether the type of the \emph{source\_variable} type and the \emph{destination\_type} are compatible with the inheritance relationship through the class hierarchy. \emph{dynamic\_cast} is used in the form of \emph{dynamic\_cast<destination\_type> (source\_variable)}. \emph{dynamic\_cast} is a typecast operation that verifies compatibility at runtime. At the time of typecasting, the type conversion compatibility between the actual type of the object stored in the \emph{source\_variable} and the \emph{destination\_type} is verified. To verify this type conversion compatibility, \emph{dynamic\_cast} searches RTTI information of an actual object stored in \emph{source\_variable}. reinterpret\_cast and static\_cast typecast operations exist only in the source code and these are used to verify type compatibility at compile time, and the typecast operator information is not left in the compiled binary. On the other hand, when \emph{dynamic\_cast} typecast operation is used, additional code for performing type verification at runtime is inserted into the binary, so information about the typecast operation remains in the compiled binary. C-style cast is used in the form of \emph{(destination\_type)source\_variable}. When C-style cast is used, the compiler tries typecast in the order of \emph{const\_cast} (typecast operation to remove const of variables), \emph{static\_cast}, and \emph{reinterpret\_cast} and uses the first successful typecast. That is, \emph{C-style} cast is the same as one of \emph{static\_cast} and \emph{reinterpret\_cast}.}\newline
            \emph{The type confusion bug occurs when a variable is type-casted to an incompatible destination type. Typecast operations, except \emph{dynamic\_cast}, verify type compatibility only at compile time. Therefore, if a source variable that is not compatible with the destination type is provided at runtime, the compatibility of the type for type conversion cannot be verified. As a result, the program is executed with the variable being considered as the wrong type, leading to unintended behavior. Although it is possible to prevent incorrect type conversion by using \emph{dynamic\_cast} operation that verifies type compatibility at runtime, \emph{dynamic\_cast} searches for RTTI to verify type conversion compatibility, which leads to a decrease in program performance. Therefore, large-scale software [16][11] does not use \emph{dynamic\_cast} for most type conversions. As a result, the compatibility of typecasting can be verified only at compile time and there is a possibility of a type confusion bug.''}  
        \end{itemize}\\ 

    \multirow{5}{*}{\parbox[c][0.6\textheight][c]{2cm}{\centering\rotatebox{90}{\textsc{T~~y~~p~~e~~~~~~C~~o~~n~~f~~u~~s~~i~~o~~n}}}}

        &\footnotesize{\cite{fan2023accelerating}} &
        \begin{itemize}
            \item\footnotesize\emph{``Our approach exclusively focuses on detecting type confusion vulnerabilities. Following previous works on type confusion bugs detection [ 17, 24, 27, 30, 35, 44], we assume the attackers are able to perform arbitrary memory reads, but arbitrary writes are prohibited. We assume that orthogonal defense mechanisms are deployed for other vulnerabilities such as integer overflow and memory safety.''}
            \item\footnotesize\emph{``LLVM-CFI reports two type confusion bugs (sectionbasedconfig.cc:1023, sectionbasedconfig.cc:1068) in the benchmark omnetpp\_r from SPEC CPU 2017. The two type confusion bugs are caused by casting an object of a parent class (\texttt{KeyValue1}) to an object of a child class (\texttt{KeyValue2}). Then the downcasted object is used to access virtual method \texttt{getKey()} and field value. However, both the virtual method \texttt{getKey()} and the field value are defined in class \texttt{KeyValue1} and inherited by class \texttt{KeyValue2}, which means the two accesses using the downcasted object are benign accesses. \textsc{htade} can successfully identify these two safe downcast instructions and remove the redundant runtime verifications before them. Thereby preventing these two false alarms.''}  
        \end{itemize}\\ 

        &\cellcolor{light-gray}\footnotesize{\cite{farkhani2018effectiveness}} &
        \cellcolor{light-gray}\begin{itemize}
            \item\footnotesize\emph{``The challenges in creating a practical CFI has resulted in the development of a new type of CFI based on runtime type checking (RTC). RTC-based CFI has been implemented in a number of recent practical efforts such as GRSecurity Reuse Attack Protector (RAP) and LLVM-CFI.''}
            \item\footnotesize\emph{``To the best of our knowledge, three implementations of RTC are available that protect both forward and backward edges with type checking: KCFI [23], RAP [46], and MCFI [44]. Other approaches such as IFCC [47], LLVM-CFI [21], and TypeArmor [49] only protect the forward edge. [...] RAP and LLVM-CFI also provide the most stable implementations of RTC as they are targeting production environments, and are not research prototypes (RAP has even been applied to the Linux distribution, Subgraph OS [41]). In addition, RAP provides a more accurate CFG than LLVM-CFI because it removes static functions from the target set, unlike LLVM-CFI. Furthermore, neither RAP nor LLVM-CFI limit the target set to address-taken functions.''}  
        \end{itemize}\\
        
        &\footnotesize{\cite{jeon2017hextype}} &
        \begin{itemize}
            \item\footnotesize\emph{``Control-Flow Integrity (CFI) techniques [20, 34– 36] verify all indirect control-flow transfers within a program to detect control-flow hijacking. However, these techniques address the type confusion problem only partially if control flow is hijacked, i.e., they detect usage of the corrupted vtable pointer, ignoring any preceding data corruption. Similarly, vtable protection schemes [14 , 38] protect virtual calls from vtable hijacking attacks but do not block type confusion attacks. Memory safety mechanisms [24, 26 , 32] protect against spatial and temporal memory safety violations but incur prohibitively high overhead in practice. [...]  Control-flow hijacking protection and memory safety are therefore orthogonal to type confusion detection. Type confusion may be used to cause a memory safety violation. Detecting type confusion allows earlier detection of security violations for these cases.''}
            \item\footnotesize\emph{``Several Control-Flow Integrity (CFI) techniques [2, 3 , 8, 12 , 34 ,40] ensure the integrity by checking any invalid control-flow transfer within the program. However, these techniques address the type confusion problem only partially if control-flow hijacking is performed via type exploitation. Similarly, defenses [17 , 31 , 38 , 39] that protect virtual calls from vtable hijacking attacks considers only the type of the virtual calls. These schemes do not address the overall bad casting problems.''}  
        \end{itemize}\\ 
        
        &\cellcolor{light-gray}\footnotesize{\cite{li2020finding}} &
        \cellcolor{light-gray}\begin{itemize}
            \item\footnotesize\emph{``\emph{Type confusion.} Polymorphic objects could be used as base class instances (i.e., up-cast) or derived class instances (i.e., down-cast) at runtime. Unsafe down casts will wrongly interpret a smaller base class object as a larger derived class object, causing memory corruption when such an object is dereferenced.''}
            \item\footnotesize\emph{``\textbf{Table 2: Effectiveness of CFI mechanisms against typical attack scenarios. The left-most column represents the specific attack scenario, and the right-most column represents the detail type of attack. The meaning of each symbol is shown in the label below the table.}''}  
        \end{itemize}\\ 

        &\footnotesize{\cite{typeconf}} &
        \begin{itemize}
            \item\footnotesize\emph{``When a memory buffer is accessed using the wrong type, it could read or write memory out of the bounds of the buffer, if the allocated buffer is smaller than the type that the code is attempting to access, leading to a crash and possibly code execution.''}
            \item\footnotesize\emph{``ChildOf 
            \begin{itemize}
                \item Class-704
                \item Incorrect Type Conversion or Cast''
            \end{itemize}}
            \item\footnotesize\emph{``CanPrecede  
            \begin{itemize}
                \item Class-119
                \item Improper Restriction of Operations within the Bounds of a Memory Buffer''
            \end{itemize}} 
            \item\footnotesize\emph{``Languages: 
            \begin{itemize}
                \item C (Undetermined Prevalence)
                \item C++ (Undetermined Prevalence)''
            \end{itemize}}  
        \end{itemize}\\ 

    \multirow{2}{*}{\parbox[c][0.3\textheight][c]{2cm}{\centering\rotatebox{90}{\textsc{T~~y~~p~~e~~~~~~C~~o~~n~~f~~u~~s~~i~~o~~n}}}}
        
        &\cellcolor{light-gray}\footnotesize{\cite{muntean2018castsan}} &
        \cellcolor{light-gray}\begin{itemize}
            \item\footnotesize\emph{``A typical scenario, where type confusion manifests itself, occurs when an object of a parent class is cast into a descendant class type. This is typically unsafe, if the parent class lacks fields expected by the descendant type object. Thus, the program may interpret the non-existent field or function in the descendant class constructor as data, or as a virtual function pointer in another context. Object type confusion leads to undefined behavior according to the C++ language draft [1]. Further, undefined behavior can lead to memory corruption, which in turn leads to exploits such as code reuse attacks (CRAs) [6] or even to advanced versions of CRAs including the COOP attack [30]. These attacks violate the control flow integrity (CFI) [2,3] of the program, by bypassing currently available OS-deployed security mechanisms such as DEP [26] and ASLR [28]. In summary, the lack of object type safety and, more broadly, memory safety can lead to object type confusion vulnerabilities (i.e., CVE-2017-3106 [12]).}
            \item\footnotesize\emph{``Table 1 depicts the currently available solutions, which can be used for C++ object type confusion detection during runtime. The tools come with the following limitations: (1) high runtime overhead (mostly due to the usage of a compiler runtime library), (2) limited type checking coverage, (3) lack of support for non-polymorphic classes, (4) absence of threads support, and (5) high maintenance overhead, as some tools require a manually maintained blacklist.''}
            \item\footnotesize\emph{``We present \textsc{CASTSAN}, a Clang/LLVM compiler-based solution, usable as an always- on sanitizer for detecting all types of polymorphic-only object type confusions during runtime, with comparable coverage to Clang-CFI [8].''}  
        \end{itemize}\\ 
        
        &\footnotesize{\cite{pang2018mapping}} &
        \begin{itemize}
            \item\footnotesize\emph{``However, C++ suffers from memory and type safety, which leave programs with huge risks. Among them, type confusion is one of the most common and dangerous vulnerabilities in C++, as it mixes the above two safety problems.''}
            \item\footnotesize\emph{``Type confusion occurs when the program intercepts an object of a type as an object of another different type in a process of unsafe typecasting. It is common that a program casts an object of a parent class to a descendant class. Thus, there exists a possibility of unsafe typecasting because the parent class may lack some fields or virtual functions of the descendant class. If the program subsequently uses such fields or functions, it may use data, say, as a regular field in one context or as a virtual function table (vtable) pointer in another. When attackers abuse a type confusion vulnerability, they can easily perform arbitrary memory read/write, sometimes even gain a higher privilege from it.}\newline 
            \emph{The statistics of CVE show a wide spread of type confusion vulnerabilities in popular applications, including Google Chrome (CVE-2017-5116), Adobe Flash Player (CVE-2018-4944), Foxit PDF Reader (CVE-2018-9942). Those exploitable type confusion bugs exactly do harm to software safety, and need to be addressed right away.''}
            \item\footnotesize\emph{``Checking typecastings at runtime is the current method to solve type confusion problems. If we choose to identify objects through existing fields embedded in the objects such as vtable pointers [18 ,29 , 38, 43], we will definitely lose control of non-polymorphic objects which have no vtable pointers, and have to blacklist the unsupported classes manually in case of an unexpected crashes. Therefore, another way of leveraging disjoint metadata [19, 22, 25] to identify objects seems better, and is widely-used in recent researches.''}
            \item\footnotesize\emph{``[2 , 5, 9, 15 , 39, 40 , 44] introduce several Control-Flow Integrity (CFI) techniques which check control-flow transfers are valid or not in the program. To a certain extent, these works can deal with the type confusion caused by control-flow hijacking.''}  
        \end{itemize}\\ 

    \multirow{3}{*}{\parbox[c][0.24\textheight][c]{2cm}{\centering\rotatebox{90}{\textsc{T~~y~~p~~e~~~~~~C~~o~~n~~f~~u~~s~~i~~o~~n}}}}

        &\cellcolor{light-gray}\footnotesize{\cite{LLVMdoc}} &
        \cellcolor{light-gray}\footnotesize {EVERYTHING} \\ 

        &\footnotesize{\cite{tice2014enforcing}} &
        \begin{itemize}
            \item\footnotesize\emph{``IFCC can support many kinds of function-pointer sets, each with different levels of precision. For example, the most precise version would have one function-pointer set per function type signature. However, real world code does not always respect the type of function pointers, so this can fail for function-pointer casts. We will focus on two simple ways of constructing function-pointer sets: (1) \emph{Single} puts all the functions into a single set; and (2) \emph{Arity} assigns functions to a set according to the number of arguments passed to the indirect call at the call site (ignoring the return value).''} 
        \end{itemize}\\ 
        
        &\cellcolor{light-gray}\footnotesize{\cite{zou2019tcd}} &
        \cellcolor{light-gray}\begin{itemize}
            \item\footnotesize\emph{``\textbf{Control Flow Integrity.} Type confusion bugs may lead to control flow attacks, which can be mitigated by control flow integrity [41]–[44]. CFI states that program execution must follow the control flow graph (CFG) generated at compile time. Two main challenges remain: how to make control-flow targets in the CFG precise and how to make dynamic checks at these control-flow targets efficient. In general, CFI defense mechanisms only protect code pointers. However, type confusion bugs can be exploited to corrupt not only code pointers but also other sensitive data.''}  
        \end{itemize}\\ 

\end{supertabular}
\end{center}
\twocolumn

\end{document}